\documentclass[preprint,superscriptaddress,nofootinbib]{revtex4}
\usepackage{graphicx}% Include figure files
\usepackage{dcolumn}% Align table columns on decimal point
\usepackage{bm}% bold math
\usepackage{amsmath}
\usepackage{amsfonts} % for kokuban-shotai
\usepackage{latexsym}
\usepackage{bbm}
\usepackage{color}
\usepackage{amssymb}
\usepackage{amsthm}
\usepackage{epsf}
\usepackage{epsfig}
\usepackage{caption3}
\usepackage{appendix}
\usepackage{cases}
\usepackage{subfigure}
\usepackage{multirow}
\makeatletter

\newcommand{\Rmnum}[1]{\expandafter\@slowromancap\romannumeral #1@}
\makeatother

\begin{document}

\title{Gravitational wave and collider signals in complex two-Higgs doublet model with dynamical CP-violation at finite temperature}

\author{Xiao Wang}%
%\email{wangxiao2016@ihep.ac.cn}
\affiliation{Theoretical Physics Division, Institute of High Energy Physics, Chinese Academy of Sciences, 19B Yuquan Road, Shijingshan District, Beijing 100049, China}
\affiliation{School of Physics, University of Chinese Academy of Sciences, Beijing 100049, China}

\author{Fa Peng Huang}%
%\email{fapeng.huang@wustl.edu}
\affiliation{Department of Physics and McDonnell Center for the Space Sciences, Washington University, St.
Louis, MO 63130, USA}
	
\author{Xinmin Zhang}
\affiliation{Theoretical Physics Division, Institute of High Energy Physics, Chinese Academy of Sciences, 19B Yuquan Road, Shijingshan District, Beijing 100049, China}
\affiliation{University of Chinese Academy of Sciences, Beijing, China}

\bigskip
	
\date{\today}
	
\begin{abstract}
Extra CP-violating source for electroweak baryogenesis can dynamically appear at finite temperature in the complex two-Higgs doublet	model, which might help to alleviate the strong constraints from the electric dipole moment experiments. In this scenario, we study the detailed phase transition dynamics and the corresponding gravitational wave signals in synergy with the collider signals at future lepton colliders. For some parameter spaces, various phase transition patterns can occur, such as the multi-step phase transition and supercooling. Gravitational waves  complementary to collider signals can help to pin down the underlying phase transition dynamics or different phase transition patterns.		
\end{abstract}
	

\maketitle
	
%\tableofcontents%temporal
%%%%%%%%%%%%%%%%%%%%%%%%%%%%%%%%%%%%%%%%%%%%%
%								Introduction
%%%%%%%%%%%%%%%%%%%%%%%%%%%%%%%%%%%%%%%%%%%%%
\section{Introduction}
After the observation of the gravitational wave (GW) by the Advanced Laser Interferometer Gravitational Wave Observatory~\cite{Abbott:2016blz}, a new era of GW astronomy has been initiated and the GW detector provides a new technique to study the fundamental physics. Especially, electroweak (EW) baryogenesis~\cite{Kuzmin:1985mm,Trodden:1998ym,Morrissey:2012db,White:2016nbo,Mazumdar:2018dfl}, which is aimed to explain the baryon asymmetry of the Universe, becomes a promising and testable mechanism after the discovery of GW and Higgs boson. To generate the observed baryon asymmetry of the Universe, all three Sakharov conditions need to be satisfied~\cite{Sakharov:1967dj}. These conditions are baryon number violation, C and CP violation, and the departure from the thermal equilibrium or CPT violation. An essential ingredient for a successful EW baryongenesis is the process of a strong first-order phase transition (FOPT) which can achieve the departure from thermal equilibrium. As a by product, the phase transition GW signal induced by a strong FOPT can potentially be detected by the future space-based GW interferometers.

In the standard model (SM), the discovery of Higgs boson by ATLAS \cite{Aad:2012tfa} and CMS \cite{Chatrchyan:2012xdj} shows
that a strong FOPT can not be generated for a 125 GeV Higgs boson based on lattice simulation \cite{Kajantie:1996mn,Kajantie:1996qd}.
It is just a smooth crossover for 125 GeV Higgs boson in the SM.
The CP violation source is also too weak for successful EW baryogenesis in the SM.
Thus, the extension of the SM is needed to provide a strong FOPT and a large enough CP violation for successful EW baryogenesis.
One of the simplest extension of the SM, which is the so-called 2-Higgs doublet model (2HDM), is the SM with an additional $SU(2)_L$ scaler doublet, where the sphaleron process was studied in Ref.~\cite{Kastening:1991nw}.
However, current electric dipole moments (EDM) experiments~\cite{Andreev:2018ayy} have put strong constraints on the CP-violating source at zero temperature for most of the new physics models.
In this work, we focus on the complex 2HDM (C2HDM).
Recent study~\cite{Basler:2017uxn} has shown that there are viable parameter spaces in the C2HDM which can produce a strong FOPT with spontaneous CP violation based on the criterion $v_c/T_c >1$, where $v_c$ is the vacuum expectation value (VEV) at the critical temperature $T_c$.
They also discuss the collider phenomenology including the modification of Higgs trilinear coupling and Higgs boson pair production at hadron collider.
Further, Ref.~\cite{Fontes:2017zfn} has revisited the constraints from colliders and EDM, and future predictions in details.
Based on these two comprehensive studies~\cite{Basler:2017uxn,Fontes:2017zfn},
we investigate the phase transition dynamics with different phase transition patterns.
Besides the dynamical CP-violating behavior, we also find the multi-step phase transition patterns and supercooling patterns.
The dynamical process might help to provide the CP-violating source for successful EW baryogenesis
\footnote{Refs.~\cite{Cline:2011mm,Dorsch:2016nrg} studied the realization of EW baryogenesis in the CP-violating 2HDM.  Although, this work is motivated by EW baryogenesis, we do not study the realization of EW baryogenesis in this work. The GW signal in the CP-violating 2HDM with CP-violating vacuum at zero temperature was also studied in Ref.~\cite{Dorsch:2016nrg}. However, for the model used in this work, the vacuum is real at zero temperature.}.
We discuss other possible approaches to explore this scenario in C2HDM.
On one hand, during a strong FOPT, detectable GWs can be produced by three mechanisms: bubble collisions, sound waves, and magnetohydrodynamic turbulence.
Based on the viable parameters from Refs.~\cite{Basler:2017uxn,Fontes:2017zfn}, we discuss the possibility to detect the GW signals by the future space-based experiments, such as the approved Laser Interferometer Space Antenna (LISA) \cite{lisa:2017} (launch in 2034 or even earlier), Deci-hertz Interferometer Gravitational wave Observatory (DECIGO) \cite{Seto:2001qf,Kawamura:2011zz}, Ultimate-DECIGO (U-DECIGO) \cite{Kudoh:2005as}, Big Bang Observer (BBO) \cite{Corbin:2005ny}, Taiji \cite{Hu:2017mde,Guo:2018npi}, and TianQin \cite{Luo:2015ght,Hu:2018yqb}.
The dynamical CP-violation behavior can escape the strong constraints from electric dipole moment (EDM) measurements~\cite{Baldes:2016rqn,Baldes:2016gaf,Bruggisser:2017lhc,Bruggisser:2018mus,Chala:2018opy,Ellis:2019flb}.
On the other hand, the strong FOPT could obviously modify the Higgs trilinear coupling and thus can be tested by the future lepton collider, such as Circular Electron-Positron Collider
(CEPC)~\cite{CEPCStudyGroup:2018ghi},
 International Linear Collider (ILC)~\cite{Fujii:2017vwa} as well as
Future Circular Collider (FCC-ee)~\cite{Abada:2019zxq}. Combined with the GW signals, they can make a complementary test on this scenario and the underlying phase transition patterns.

This paper is organized as follows. In Section II, we describe the C2HDM and the basic idea of dynamical CP-violation at finite temperature.
In section III, the one-loop effective potential at finite temperature and the renormalization prescription are presented.\footnote{In Appendix A, we present the thermal corrections of the masses for the C2HDM in the Landau gauge. In Appendix B, we derive the field dependent mass matrix elements
for the gauge bosons, the scalar bosons and the top quark for C2HDM in the Landau gauge.}
In section IV, we investigate the phase transition dynamics including the corresponding GW signals and its correlation with the collider signatures.
We discuss the consistent check of the dynamical CP-violation and supercooling case in section V.
Section VI contains our conclusions.

\section{Model with dynamical CP-violation}
The tree-level potential of the C2HDM can be written as
\begin{align}
\begin{split}
V_{\text{tree}} &= m_{11}^2 \Phi_1^\dagger \Phi_1 + m_{22}^2
\Phi_2^\dagger \Phi_2 - \left[m_{12}^2 \Phi_1^\dagger \Phi_2 +
\mathrm{h.c.} \right] + \frac{1}{2} \lambda_1 ( \Phi_1^\dagger
\Phi_1)^2 +\frac{1}{2} \lambda_2 (\Phi_2^\dagger \Phi_2)^2 \\
&\quad + \lambda_3 (\Phi_1^\dagger \Phi_1)(\Phi_2^\dagger\Phi_2) +
\lambda_4 (\Phi_1^\dagger \Phi_2)(\Phi_2^\dagger \Phi_1)
+ \left[ \frac{1}{2} \lambda_5 (\Phi_1^\dagger\Phi_2)^2  +
\mathrm{h.c.} \right] \; ,
\end{split}\label{eq:treepot}
\end{align}
where $m_{12}^2$ and $\lambda_5$ are complex numbers.
% and $\text{arg}(\lambda_5)\neq 2\text{arg}(m_{12}^2)$.
It is obvious that the C2HDM has a softly broken $Z_2$ symmetry ($\Phi_1\to\Phi_1,\Phi_2\to-\Phi_2$).
At zero temperature, we have
\begin{equation}\label{vv0}
\Phi_1^{(0)}= \frac{1}{\sqrt{2}} \left(\begin{array}{c} \rho_1 + i\eta_1\\
v_{1} + \zeta_1 + i\psi_1
\end{array}\right)
\quad
\Phi_2^{(0)}= \frac{1}{\sqrt{2}} \left(\begin{array}{c} \rho_2 + i\eta_2\\
v_{2}  + \zeta_2 + i\psi_2
\end{array}\right).
\end{equation}
We have real vacuum at zero temperature.
This model has been extensively studied including the EDM constraints and collider phenomenology, such as the recent works~\cite{Basler:2017uxn,Fontes:2017zfn} and references therein.
However, at finite temperature,
there would be dynamical CP-violating behavior~\cite{Basler:2017uxn} as
\begin{equation}\label{sp}
\Phi_1^{(T)} = \frac{1}{\sqrt{2}} \left(\begin{array}{c} \rho_1 + i\eta_1\\
\tilde{v}_{1} + \zeta_1 + i\psi_1
\end{array}\right)
\quad
\Phi_2^{(T)} = \frac{1}{\sqrt{2}} \left(\begin{array}{c} \tilde{v}_{CB} + \rho_2 + i\eta_2\\
\tilde{v}_{2} + i\tilde{v}_{CP} + \zeta_2 + i\psi_2
\end{array}\right).
\end{equation}
The $\tilde{v}$ with tilde represents the VEV at finite temperature.
This is the starting point of this work. It means there exists extra CP-violation at high temperature, which might provide the CP-violating source for successful EW baryogenesis.
At zero temperature, this extra CP-violating source disappears to escape the severe EDM constraints. To consider more general situation, we also assume there is charge-breaking at high temperature.
In Section V, we show the numerical results on the evolution of these CP-violating source with the decreasing of the temperature which confirms the starting point is consistent. Various phase transition patterns can also be triggered based on Eq.~(\ref{sp}), which are discussed carefully in next section.

For more compact form, the VEVs at zero temperature are denoted as
\begin{equation}
\tilde{v}_{1}(T=0)=v_1,  \quad \tilde{v}_{2}(T=0) =  v_2,\quad \tilde{v}_{CP}(T=0)=v_{CP} =0,\quad \tilde{v}_{CB} (T=0)= v_{CB} = 0,
\end{equation}
with this convention,
\begin{equation}
v \equiv \sqrt{v_1^2 + v_2^2 + v_{CP}^2 + v_{CB}^2} = \sqrt{v_1^2 + v_2^2},
\end{equation}
where $v\approx246$ GeV is the SM VEV, and the stationary conditions are
\begin{equation}
\frac{\partial V_{tree}}{\partial \Phi_i}\Bigg|_{\Phi_i = \langle\Phi_i\rangle} = 0,\quad
\frac{\partial V_{tree}}{\partial \Phi_i\dagger}\Bigg|_{\Phi_i = \langle\Phi_i\rangle} = 0, \quad i=1,2,
\end{equation}
and give the following relations
\begin{equation}
m_{11}^2 = Re(m_{12}^2)\frac{v_2}{v_1} - \frac{v_1^2}{2}\lambda_1 - \frac{v_2^2}{2}\lambda_{345},
\end{equation}
\begin{equation}
m_{22}^2 = Re(m_{12}^2)\frac{v_1}{v_2} - \frac{v_2^2}{2}\lambda_2 - \frac{v_1^2}{2}\lambda_{345},
\end{equation}
\begin{equation}
 \frac{v_1v_2Im(\lambda_5)}{2}=Im(m_{12}^2) ,
\end{equation}
where
\begin{equation}
 \lambda_3 + \lambda_4 + Re(\lambda_5) \equiv \lambda_{345} .
\end{equation}
We introduce a mixing angle $\Theta$, which is defined as
\begin{equation}
\tan\Theta = \frac{v_2}{v_1},
\end{equation}
then transform the fields into a new basis
\begin{equation}
\zeta_3 = -\sin\Theta\psi_1 + \cos\Theta\psi_2, \quad
A = \cos\Theta\psi_1 + \sin\Theta\psi_2  \,\,.
\end{equation}
In this C2HDM, there is only one independent CP-violating phase,
which satisfies the relations in Eqs.~(7-9) at zero temperature.
Compared to Ref.~\cite{Dorsch:2016nrg}, $v_1,v_2$ are real in this work as shown in Eq.~(\ref{vv0}).
We can use similar invariant definition of the CP-violating phase as in Refs.~\cite{Dorsch:2016nrg,Inoue:2014nva} below
\begin{eqnarray}
  \delta_1 &=& \arg[(m_{12}^2)^2 \lambda_5^{\ast}]  \,\,, \\
  \delta_2 &=& \arg[m_{12}^2 v_1 v_2^{\ast} \lambda_5^{\ast}] \,\,,\\
  \mid m_{12}^2\mid \sin(\delta_1-\delta_2) &=& v^2 \sin \Theta \cos \Theta \mid \lambda_5 \mid \sin(\delta_1-2\delta_2)\,\,.
\end{eqnarray}
However, at finite temperature, it becomes difficult to define such simple invariant quantity, which is investigated at zero temperature~\cite{Gunion:2005ja,Branco:2011iw}.
And the extra CP violation can not be rotated away by field redefinition at high temperature while keeping the original CP-violation at zero temperature unchanged since the temperature dependent CP violation occurs in the same basis used in the zero temperature case. It is numerically confirmed in the discussion section V.A.

In the C2HDM, the neutral components $\zeta_1$, $\zeta_2$ and $\zeta_3$ mix into the neutral mass eigenstates $H_i~(i=1,2,3)$ through the mixing matrix
\begin{equation}
\left(\begin{array}{c} H_1 \\ H_2 \\H_3
\end{array}\right) = R\left(\begin{array}{c}\zeta_1 \\ \zeta_2 \\ \zeta_3
\end{array}\right).
\end{equation}
The mixing matrix $R$ can diagonalize the neutral mass matrix
\begin{equation}
M_{ij} = \frac{\partial^2V}{\partial\zeta_i\partial\zeta_j}\,\,\,,
\end{equation}
and derive
\begin{equation}
RMR^T = \text{diag}(m_1^2, m_2^2, m_3^2)\,\,\,,
\end{equation}
where $m_1 \leq m_2 \leq m_3$ are the masses of the neutral Higgs bosons. We can parametrize the matrix $R$ as the following~\cite{ElKaffas:2007rq}
\begin{equation}
R = \left(\begin{array}{ccc}
c_1c_2 & s_1c_2 & s_2 \\
- c_1s_2s_3 - s_1c_3 & c_1c_3 - s_1s_2s_3 & c_2s_3 \\
-c_1s_2c_3 + s_1s_3 & - c_1s_3 - s_1s_2c_3 & c_2c_3
\end{array}\right),
\end{equation}
where $s_i = \sin\theta_i$, $c_i = \cos\theta_i (i=1,2,3)$, and $-\frac{\pi}{2} \le \theta_i < \frac{\pi}{2}$~\cite{Basler:2017uxn,Fontes:2017zfn}.
%\begin{equation}
%-\frac{\pi}{2} \le \alpha_i < \frac{\pi}{2}\,\,\,.
%\end{equation}
Note the above mixing matrix is valid at zero temperature. When we consider the finite-temperature situation in the next section, this result should be modified. The Higgs potential in Eq.\eqref{eq:treepot} has 9 independent parameters. We follow Ref.~\cite{ElKaffas:2006gdt} and choose 9 input parameters $v$, $\tan\Theta$, $m_{H_{\pm}}$, $\theta_1$, $\theta_2$, $\theta_3$, $m_1$, $m_2$, and $Re(m_{12}^2)$. For these input parameters, $m_3$ can be expressed as
\begin{equation}
m_3^2 = \frac{m_1^2R_{13}(R_{12}\tan\Theta - R_{11}) + m_2^2R_{23}(R_{22}\tan\Theta - R_{21})}{R_{33}(R_{31} - R_{32}\tan\Theta)}\,\,\,.
\end{equation}
The analytic relations between the above parameter set and the coupling parameters $\lambda_i$ in the original Lagrangian can be written as \cite{Fontes:2014xva}
\begin{equation}
\lambda_1 = \frac{1}{v^2\cos^2\Theta}\left[m_1^2c_1^2c_2^2 + m_2^2(c_3s_1 + c_1s_2s_3)^2 + m_3^2(c_1c_3s_2 - s_1s_3)^2 - \mu^2\sin^2\Theta\right],\notag
\end{equation}
\begin{equation}
\lambda_2 = \frac{1}{v^2\sin^2\Theta}\left[m_1^2s_1^2c_2^2 + m_2^2(c_1c_3 - s_1s_2s_3)^2 + m_3^2(c_3s_1s_2 + c_1s_3)^2 - \mu^2\cos^2\Theta\right], \notag
\end{equation}
\begin{equation}
\begin{split}
\lambda_3 &= \frac{1}{v^2\sin\Theta\cos\Theta}[(m_1^2c_2^2 + m_2^2(s_2^2s_3^2 - c_3^2) + m_3^2(s_2^2c_3^2 - s_3^2))c_1s_1\\
&\quad+ (m_3^2 - m_2^2)(c_1^2 - s_1^2)s_2c_3s_3 ]- \frac{\mu^2 - 2m_{H_{\pm}}^2}{v^2}
\end{split}, \notag
\end{equation}
\begin{equation}
\lambda_4 = \frac{m_1^2s_2^2 + (m_2^2s_3^2 + m_3^2c_3^2)c_2^2 + \mu^2 - 2m_{H_{\pm}}^2}{v^2}, \notag
\end{equation}
\begin{equation}
Re(\lambda_5) = \frac{-m_1^2s_2^2 - (m_2^2s_3^2 + m_3^2c_3^2)c_2^2 + \mu^2}{v^2}, \notag
\end{equation}
\begin{equation}
Im(\lambda_5) = \frac{2c_2}{v^2\sin\Theta}\left[(-m_1^2 + m_2^2s_3^2 + m_3^2c_3^2)c_1s_2 + (m_2^2 - m_3^2)s_1s_3c_3\right],
\end{equation}
where
\begin{equation}
\mu^2 = \frac{v^2}{v_1^2v_2^2}Re(m_{12}^2)\,\,\,.
\end{equation}
In general, 2HDM can be classified into type \Rmnum{1}, type \Rmnum{2}, lepton-specific and flipped, according to the interactions of the fermions to the Higgs doublets. In this work we only study type \Rmnum{1} case as an example, and only consider the top quark's contribution to the EW phase transition among all the fermions.
For EW phase transition, there exist only slight differences when switching to a different type of 2HDM  since only top Yukawa coupling is considered among all the fermions in the phase transition calculations. However, for the constraints from colliders and EDM, different types have different constraints, such as the bound on the $m_{H_{\pm}}-\tan \Theta$ plane. For example, the constraint from EDM in type I model is  weaker than the one in type II~\cite{Fontes:2017zfn,Inoue:2014nva}. There is larger parameter space for type I 2HDM.

\section{Phase transition dynamics and CP-violation at finite temperature}
To study the phase transition dynamics in the C2HDM, we use the finite-temperature effective field theory \cite{Coleman:1973jx,Dolan:1973qd,Quiros:1999jp}. The full one-loop finite-temperature effective potential reads
\begin{equation}
V_{eff}(\tilde{v}, T) \equiv V_{tree}(\tilde{v}) + V_{CW}(\tilde{v}) + V_{CT}(\tilde{v}) + V_T(\tilde{v}, T)\,\,\,, \label{eq.effpot}
\end{equation}
where $V_{tree}$, which is obtained by replacing the doublets with their classical background fields ($\tilde{v}_1, \tilde{v}_2, \tilde{v}_{CP}, \tilde{v}_{CB}$) from Eq.~(\ref{sp}), is the tree-level potential at zero temperature as shown in the following
\begin{align}
\begin{split}
V_{tree}(\tilde{v}) &= \frac{1}{2} m_{11}^2 \tilde{v}_{1}^2 + \frac{1}{2} m_{22}^2\left(\tilde{v}_{2}^2 + \tilde{v}_{CB}^2 + \tilde{v}_{CP}^2\right) - Re(m_{12}^2)\tilde{v}_{1}\tilde{v}_{2} + Im(m_{12}^2)\tilde{v}_{1}\tilde{v}_{CP} + \frac{1}{8}\lambda_1\tilde{v}_{1}^4 \\
&\quad+ \frac{1}{8}\lambda_2\left(\tilde{v}_{2}^2 + \tilde{v}_{CP}^2 + \tilde{v}_{CB}^2\right)^2 +   \frac{1}{4}\lambda_3\tilde{v}_{1}^2\left(\tilde{v}_{2}^2 + \tilde{v}_{CB}^2 + \tilde{v}_{CP}^2\right) + \frac{1}{4}\lambda_4\tilde{v}_{1}^2\left(\tilde{v}_{2}^2 + \tilde{v}_{CP}^2\right) \\
&\quad+ \frac{1}{4}Re(\lambda_5)\tilde{v}_{1}^2(\tilde{v}_{2}^2 - \tilde{v}_{CP}^2) - \frac{1}{2}Im(\lambda_5)\tilde{v}_{1}^2\tilde{v}_2\tilde{v}_{CP}\,\,\,.
\end{split}
\end{align}
$V_{CW}$ is the Coleman-Weinberg potential (CW) at zero temperature. In the $\overline{\text{MS}}$ scheme, the CW potential
can be written as
\begin{equation}
V_{CW}(\tilde{v}) = \frac{1}{64\pi^2} \sum_{s} n_s m_s^4(\tilde{v})\left[\log{\frac{m_s^2(\tilde{v})}{\mu^2}} - C_s\right],
\end{equation}
where $\tilde{v} \equiv \{\tilde{v}_{1}, \tilde{v}_{2}, \tilde{v}_{CP}, \tilde{v}_{CB}\}$, and $m_{s}^2(\tilde{v})$ is the eigenvalue for the particle $s$ in the mass matrix in terms of the background fields $\tilde{v}$. The details are shown in Appendix~\ref{B}. $n_s$ denotes the numbers of the degree of freedom. Because of the charge-breaking VEV, photon becomes massive roughly from several tens GeV to about 20 GeV during the phase transition. It mainly contributes to the phase transition at high temperature and quickly becomes massless when the temperature becomes lower. There would be no other important cosmological effects except for the EW phase transition.
And we have to take into account different masses and numbers of degree of freedom for the charge conjugated particles.
For each particle $s$, the numbers of degree of freedom are $\{n_{H_i}, n_{A}, n_{H^+}, n_{H^-}, n_{G^+}, n_{G^-}, n_{W^+}, n_{W^-}, n_Z, n_{\gamma}, n_{t}, n_{\bar{t}}\} = \{1, 1, 1, 1, 1, 1,3,3,3,3,-6,-6\}$
and the constants $C_s$ are
\begin{equation}
C_s = \begin{cases}
\frac{5}{6},& s=W^{\pm},Z,\gamma\\
\frac{3}{2},& \text{others}
\end{cases}\,\,.
\end{equation}
The masses and the mixing angles with one-loop corrections are different from those extracted from the tree-level potential. To enforce the one-loop corrected masses and the mixing angles to be equal to the tree-level values, we use the on-shell renormalization prescription as in Refs.~\cite{Basler:2016obg,Basler:2017uxn,Basler:2018cwe}. Then, a counterterm potential $V_{CT}$ is added to the one-loop effective potential.
The general formula of the counterterm contribution $V_{CT}$ reads~\cite{Basler:2018cwe}
\begin{equation}
V_{CT} = \sum_{i=1}^{n}\frac{\partial V_{tree}}{\partial p_i}\delta p_i + \sum_{k = 1}^{m}\delta T_k(\phi_k + \tilde{v}_k),
\end{equation}
where $\delta p_i$ and $n$ are the counterterms and the number of parameters of the tree-level potential, respectively.
	$\delta T_k$ denotes the counterterms of the tadpole $T_k$, and $m$ is the number of background field or the number of field that is allowed for the development of a non-zero VEV.
In the C2HDM, the counterterm potential can be written as
\begin{align}
\begin{split}
V_{CT} & = \delta m_{11}^2\Phi_1^\dagger\Phi_1 + \delta m_{22}^2\Phi_2^\dagger\Phi_2 - [(\delta Re(m_{12}^2) + i\delta Im(m_{12}^2))\Phi_1^\dagger\Phi_2 + \mathrm{h.c.}] \\
&\quad+ \frac{1}{2}\delta\lambda_1(\Phi_1^\dagger\Phi_1)^2 + \frac{1}{2}\delta\lambda_2(\Phi_2^\dagger\Phi_2)^2 + \delta\lambda_3(\Phi_1^\dagger\Phi_1)(\Phi_2^\dagger\Phi_2) + \delta\lambda_4(\Phi_1^\dagger\Phi_2)(\Phi_2^\dagger\Phi_1)\\
&\quad+ \frac{1}{2}[(\delta Re(\lambda_5) + i\delta Im(\lambda_5))(\Phi_1^\dagger\Phi_2)^2 + \mathrm{h.c.}]\\
&\quad+ \delta T_1(\zeta_1 + \tilde{v}_{1}) + \delta T_2(\zeta_2 + \tilde{v}_2) + \delta T_{CP}(\psi_2 + \tilde{v}_{CP}) + \delta T_{CB}(\rho_2 + \tilde{v}_{CB}).
\end{split}
\end{align}
The on-shell renormalization conditions at zero temperature are
\begin{gather}
\partial_{\phi_i}V_{CW}(\phi)\Big|_{\phi=\langle\phi^c\rangle_{T=0}}+ \partial_{\phi_i}V_{CT}(\phi)\Big|_{\phi=\langle\phi^c\rangle_{T=0}}=0, \notag \\
\partial_{\phi_i}\partial_{\phi_j}V_{CW}(\phi)\Big|_{\phi=\langle\phi^c\rangle_{T=0}}+\partial_{\phi_i}\partial_{\phi_j}V_{CT}(\phi)\Big|_{\phi=\langle\phi^c\rangle_{T=0}}=0,
\end{gather}
where
\begin{equation}
\phi_i \equiv \{\rho_1,\eta_1, \rho_2, \eta_2, \zeta_1, \psi_1, \zeta_2, \psi_2 \} \,\,,
\end{equation}
\begin{equation}
\langle\phi^c\rangle_{T=0}= \{0, 0, 0, 0, v_1, 0, v_2, 0\}\,\,.
\end{equation}
The second derivatives of the CW potential lead to the well-known problem of infrared (IR) divergences for the Goldstone bosons \cite{Camargo-Molina:2016moz,Martin:2014bca,Elias-Miro:2014pca,Casas:1994us} in the Landau gauge.
In practice, we can introduce an IR regulator for the Goldstones and then discard the terms proportional to the IR divergence.
Previous study \cite{Camargo-Molina:2016moz} has dealt with this problem and derived analytic formulas for the first and second derivatives of the CW potential in the physical basis
\begin{align}
\begin{split}
 \partial_{\phi_i}V_{CW}(\phi)\Big|_{\phi=\langle\phi^c\rangle_{T=0}} = O_{ij}^{H}\sum_s\frac{(-1)^{\chi_s}(1 + \chi_s)}{32\pi^2}m_{(s)a}^2\lambda_{(s)aaj}\left(\log{\frac{m_{(s)a}^2}{\mu^2}} - C_{s} + \frac{1}{2}\right),
\end{split}
\end{align}
\begin{align}
\begin{split}
 \partial_{\phi_i}\partial_{\phi_j}V_{CW}(\phi)\Big|_{\phi=\langle\phi^c\rangle_{T=0}}
&= O_{ik}^{H}O_{jl}^{H}\sum_s\frac{(-1)^{\chi_s}(1 + \chi_s)}{32\pi^2}S_{ij}\Bigg[\lambda_{(s)abj}\lambda_{(s)baj}\left(f_{(s)ab}^{(1)} - C_s + \frac{1}{2}\right) \\
&\quad+ \lambda_{(s)aaij}m_{(s)a}^2\left(\log{\frac{m_{(s)a}^2}{\mu^2}} - C_s + \frac{1}{2}\right)\Bigg],
\end{split}
\end{align}
with
\begin{equation}
f_{(s)a_1a_2}^{(1)} = \sum_{x=1}^{2}\frac{m_{(s)a_x}^2\log{\frac{m_{(s)a_x}^2}{\mu^2}}}{\prod_{y \ne x}\left(m_{(s)a_x}^2 - m_{(s)a_y}^2\right)}\,\,\,,\label{f1}
\end{equation}
where $\chi_s$ is the spin of different particles, $m_{(s)a}^2$ is the physical mass of particle $s$ at zero temperature, $O_{ij}^{H}$ is the rotation matrix that transform scalar fields from Laudau gauge basis to mass eigenstate basis, $S_{ij}$ denotes symmetrization with respect to the two indexes, $\lambda_{(s)abi}$ and $\lambda_{(s)abij}$ are the cubic and quartic couplings for particle $s$ in mass eigenstate basis.
Note that we need to deal with degenerate mass limit carefully in Eq.~\eqref{f1}. For more detail, see Ref.~\cite{Camargo-Molina:2016moz}.
Then the counterterms can be expressed in terms of the derivatives of the CW potential.
For the analytic formulas of the counterterms, see Refs.~\cite{Basler:2017uxn,Basler:2018cwe}.
$V_T$ is the one-loop thermal correction including daisy resummation~\cite{Arnold:1992rz,Parwani:1991gq} at finite temperature. The thermal correction reads
\begin{equation}
V_T = \sum_{F}\frac{T^4}{2\pi^2}n_FJ_F\left(\frac{m_F^2}{T^2}\right) + \sum_{B}\frac{T^4}{2\pi^2}n_BJ_B\left(\frac{m_B^2}{T^2}\right),
\end{equation}
with the thermal functions
\begin{equation}
J_{B/F} = \int_{0}^{\infty}\mathrm{d}x x^2\log\left[1 \mp e^{-\sqrt{x^2 + m_i^2/T^2}}\right],
\end{equation}
where the plus sign is for fermions and the minus sign is for bosons, $n_F$ and $n_b$ are the degree of freedom for fermions and bosons, respectively. In order to include the contribution of daisy resummation, we make the following replacement for the scalar boson mass and the longitudinal components of the gauge boson mass
\begin{equation}
m_B^2 \to \overline{m}^2_B = m_B^2 + \Pi_B\,\,\,,
\end{equation}
where $\Pi_B$ is the thermal correction of the scalar boson and the longitudinal components of gauge boson at finite temperature, which can be found in Appendix \ref{A}.
The Debye corrected masses are applied in the all terms of $J_B$ and also used in the CW potential \cite{Parwani:1991gq}.
It is worth noticing that Parwani scheme is used in this work, while Arnold-Espinosa scheme is used in Ref.~\cite{Basler:2017uxn}.

With the full effective potential in Eq.~(\ref{eq.effpot}), we can use the method that is introduced below to numerically calculate the phase transition dynamics.
From the comprehensive studies of the C2HDM \cite{Basler:2017uxn}, we know there are viable parameter space to induce a strong FOPT.
According to the Ref.~\cite{Basler:2017uxn}, we scan the viable parameter space within the allowed parameter spaces by the current collider and EDM constraints from Ref.~\cite{Fontes:2017zfn}.

For completeness and self-consistency, we show a brief summary of the constraints from theoretical aspects, colliders and EDM data based on Ref.~\cite{Fontes:2017zfn} and references therein.
Reference~\cite{Fontes:2017zfn} has imposed all available constraints on this C2HDM and  scanned the parameter space for their phenomenological analyses. Based on their allowed parameter space, we further
limit the available parameter space to smaller parameter space to satisfy the condition of strong FOPT and the new EDM data. We partially used  the publicly available package in Ref.~\cite{Fontes:2017zfn}.
For the theoretical bounds, the perturbative unitarity and vacuum stability are considered.
Then, the constraints from the oblique corrections to EW precision observables should be taken into account
within a $2\sigma$ compatibility of the EW oblique parameters.
As for the constraints of the charged sector, the exclusion bounds on the $m_{H_{\pm}}-\tan \Theta$ plane depend
on the type of the 2HDM. An important constraint on this plane is from the measurements of $B\to X_s \gamma$. For the type I and lepton-specific models, the constraint strongly depends on $\tan \Theta$. The charged Higgs mass should be heavier than 400 GeV for  $\tan \Theta \approx 1$.
For type II and flipped models, the charged Higgs mass has to be heavier than about 580 GeV,
almost independent of $\tan \Theta$ within $2\sigma$ compatibility.
Further, the flavor constraints from $R_b\equiv\Gamma[Z\to b\bar{b}]/\Gamma[Z\to \rm hadrons]$ are considered on the $m_{H_{\pm}}-\tan \Theta$ plane.
Another important bound from Higgs boson search is checked with the \emph{HiggsBounds} code~\cite{Bechtle:2013wla}.
The \emph{C2HDM\_HDECAY} code~\cite{Fontes:2017zfn} is used to calculate all the branching ratios and decay widths of all the Higgs bosons.
There are also important constraints from various EDM constraints, such as the recent ACME data \cite{Andreev:2018ayy}. Type I is less constrained by the EDM bounds compared to type II~\cite{Inoue:2014nva}.
There are still available parameter spaces after considering all the constraints as shown in the Fig.~14 of Ref.~\cite{Fontes:2017zfn} for the new EDM data $|d_e|<1.1 \times 10^{-29}~e\cdot cm$ at $90\%$ confidence level.
In the following, we will study the collider signals at future lepton colliders. Detailed revision of the collider phenomenology and EDM is beyond the scope of this work.

Here, for simplicity, we only show 12 benchmark sets which can induce various representative phase transition patterns, and we choose $H_1$ to be the SM Higgs boson. Multi-step FOPTs, supercooling and second-order phase transition (SOPT) can occur.

In Table~\ref{tb.1}, we show 8 benchmark sets. Each parameter set can give a one-step strong FOPT, and the FOPT takes place as $(0,0,0,0)\xrightarrow{FOPT}(\tilde{v}_{1},\tilde{v}_{2},\tilde{v}_{CP},\tilde{v}_{CB})\xrightarrow{T \to  0}(v_1, v_2, 0, 0)$ with the temperature decreasing from high value to zero. Only one strong FOPT happens for these benchmark sets.

In Table.~\ref{tb.2}, two parameter sets are shown.
Each benchmark set can induce two FOPTs and they evolve as $(0,0,0,0)\xrightarrow{FOPT}(\tilde{v}_{1}^{(1)},\tilde{v}_{2}^{(1)},\tilde{v}_{CP}^{(1)},\tilde{v}_{CB}^{(1)})\xrightarrow{FOPT}(\tilde{v}_{1}^{(2)},\tilde{v}_{2}^{(2)},\tilde{v}_{CP}^{(2)},\tilde{v}_{CB}^{(2)}) \xrightarrow{T \to 0}(v_1, v_2, 0, 0)$ with the temperature decreasing from high value to zero.

Three-step phase transition can be produced for the two benchmark sets in Table~\ref{tb.3}. And they evolve like $(0,0,0,0)\xrightarrow{SOPT}(\tilde{v}_{1}^{(1)},\tilde{v}_{2}^{(1)},\tilde{v}_{CP}^{(1)},\tilde{v}_{CB}^{(1)})\xrightarrow{FOPT}(\tilde{v}_{1}^{(2)},\tilde{v}_{2}^{(2)},\tilde{v}_{CP}^{(2)},\tilde{v}_{CB}^{(2)})\xrightarrow{FOPT}(\tilde{v}_{1}^{(3)},\tilde{v}_{2}^{(3)},\tilde{v}_{CP}^{(3)},\tilde{v}_{CB}^{(3)})\xrightarrow{T \to 0}(v_1, v_2, 0, 0)$ with the temperature decreasing from high value to zero.
For these two benchmark sets, two FOPTs and one SOPT occur.

\begin{table}[htp]
	\centering
	\begin{tabular}{cccccccccc}
		\hline\hline
		& $v~\text{[GeV]}$ & $m_1~\text{[GeV]}$ & $m_2~\text{[GeV]}$ & $m_{H_{\pm}}~\text{[GeV]}$ & $Re(m_{12}^2)~[\rm GeV^2]$ & $\theta_1$ & $\theta_2$ & $\theta_3$ & $\tan{\Theta}$\\
		\hline
		$BP_1\quad$ & 246 & 125.09 & 356.779 & 581.460  & 29939 & 1.470 & 0.0223 & -0.097 & 4.17  \\
		$BP_2\quad$ & 246 & 125.09 & 603.699 & 629.564  & 73628 & 0.817 & $3.687\times10^{-3}$ & -1.557 & 1.216  \\
		$BP_3\quad$ &246 & 125.09 & 455.834 & 685.479 & 85376 & 0.880 & -0.0156 & 1.568 & 1.399 \\
		$BP_4\quad$ & 246 & 125.09 & 458.834 & 683.679 & 85376 & 0.880 & -0.0156 & 1.568 & 1.399 \\
        $BP_5\quad$ & 246 & 125.09 & 490.698 & 525.220  & 20392 & 0.932 & 0.0101 & -0.514 & 1.608  \\
		$BP_6\quad$ & 246 & 125.09 & 485.698 & 530.220  & 20392 & 0.932 & 0.0101 & -0.514 & 1.608  \\
        $BP_7\quad$ & 246 & 125.09 & 495.698 & 525.220  & 20192 & 0.932 & 0.0101 & -0.514 & 1.608  \\
        $BP_8\quad$ & 246 & 125.09 & 481.698 & 533.220  & 20192 & 0.932 & 0.0101 & -0.514 & 1.608  \\
		\hline\hline
	\end{tabular}
	%	\captionsetup{type=table}
	\caption{One-step phase transition benchmark points.}\label{tb.1}
\end{table}

\begin{table}[htp]
	\centering
	\begin{tabular}{cccccccccc}
		\hline\hline
		& $v$~[\text{GeV}]& $m_1~[\text{GeV}]$ & $m_2~[\text{GeV}]$ & $m_{H_{\pm}}~[\text{GeV}]$ & $Re(m_{12}^2)~[\rm GeV^2]$ & $\theta_1$ & $\theta_2$ & $\theta_3$ & $\tan{\Theta}$\\
		\hline
		$BP_9\quad$ & 246 & 125.09 & 430.698 & 500.220  & 20192 & 0.832 & 0.0101 & -0.514 & 1.458  \\
		$BP_{10}\quad$ & 246 & 125.09 & 440.698 & 500.220  & 20092 & 0.832 & 0.0101 & -0.514 & 1.458  \\
		\hline\hline
	\end{tabular}
	\caption{Two-step phase transition benchmark points with two FOPTs.}\label{tb.2}
\end{table}

\begin{table}[htp]
	\centering
	\begin{tabular}{cccccccccc}
		\hline\hline
		& $v~[\text{GeV}]$ & $m_1~[\text{GeV}]$ & $m_2~[\text{GeV}]$ & $m_{H_{\pm}}~[\text{GeV}]$ & $Re(m_{12}^2)~[\text{Ge}V^2]$ & $\theta_1$ & $\theta_2$ & $\theta_3$ & $\tan{\Theta}$\\
		\hline
		$BP_{11}\quad$ & 246 & 125.09 & 489.698 & 550.220  & 20392 & 0.832 & 0.0101 & -0.514 & 1.508  \\
		$BP_{12}\quad$ & 246 & 125.09 & 495.698 & 543.220  & 20292 & 0.832 & 0.0101 & -0.514 & 1.508  \\
		\hline\hline
	\end{tabular}
    \caption{Three-step phase transition benchmark points with one SOPT and two FOPTs.}\label{tb.3}
\end{table}

To obtain the parameter sets in the above Tables, we need to know the bubble dynamics during the phase transition process. The essential quantity of bubble dynamics is  the bubble nucleation rate per unit time per unit volume \cite{Linde:1980tt}
\begin{equation}
\Gamma = \Gamma_0e^{-S_E},
\end{equation}
where $S_E(T) = S_3/T$ is the Euclidean action of a critical bubble and $\Gamma_0 \propto T^4$. $S_3$ is the three-dimensional Euclidean action, which can be denoted as \cite{Linde:1980tt}
\begin{equation}
S_3 = 4\pi\int drr^2\left[\frac{1}{2}\left(\frac{d\tilde{v}_i}{dr}\right)^2 + V_{eff}(\tilde{v}_i, T)\right],
\end{equation}
where $\tilde{v}_i= \{\tilde{v}_{1}, \tilde{v}_{2}, \tilde{v}_{CB}, \tilde{v}_{CP}\}$.
To calculate the nucleation rate, we need to obtain the bubble profiles of the four scalar fields
 by solving the following bounce equations
\begin{equation}
\frac{d^2\tilde{v}_i}{dr^2} + \frac{2}{r}\frac{d\tilde{v}_i}{dr} = \frac{\partial V_{eff}}{\partial \tilde{v}_i}, \quad i = 1,2,CP,CB,
\end{equation}
with the boundary conditions
\begin{equation}
\lim_{r\to\infty}\tilde{v}_i = \tilde{v}_f, \quad \frac{d\tilde{v}_i}{dr}\Big|_{r=0} = 0,
\end{equation}
where $\tilde{v}_f$ is the false VEVs.
Conventionally, we use the so-called overshooting (undershooting) method~\cite{Coleman:1977py,Callan:1977pt} to solve the single-field bounce equation.
However, the multi-field case becomes much more complicated. We use the path deformation method, which is introduced by Ref.~\cite{Wainwright:2011kj}, to find a proper path that connects the initial and final vacuum state.
In our analysis, we make use of the publicly available package \emph{cosmoTransitons} to solve the four differential bounce equations.
Then the nucleation temperature $T_n$ is defined as the temperature at time $t_n$ at which $\Gamma$ becomes large enough to nucleate a bubble per horizon volume with the probability being $\mathcal{O}(1)$ \cite{Apreda:2001us},
\begin{equation}
\int_{0}^{t_n}dt\frac{\Gamma}{H^3} \simeq 1,
\end{equation}
where $H$ is the Hubble parameter. In other words, this condition can be simplified as
\begin{equation}
\frac{S_3(T_n)}{T_n} = 4\ln{(T_n/100 \rm GeV)} + 137.
\end{equation}
The properties of the bubbles are illustrated by two key parameters $\alpha$ and $\beta$.
Note $\alpha$ is the ratio of the latent heat $\epsilon(T_n)$ to the energy density of the radiation bath $\rho_{rad}$.
It is defined as \cite{Espinosa:2010hh}
\begin{equation}
\alpha = \frac{\epsilon(T_n)}{\rho_{rad}(T_n)},
\end{equation}
where $\rho_{rad}(T) = g_{\star}\pi^2T^4/30$, and $g_{\star}$ is the number of the relativistic degree of freedom in the thermal plasma at $T$. And $\epsilon(T_n)$ can be written as
\begin{equation}
\epsilon(T_n) = \left[-V_{eff}(\phi,T) + T\frac{\partial V_{eff}(\phi,T)}{\partial T}\right]\Bigg|_{T=T_n}.
\end{equation}
Moreover, the parameter $\beta$ is defined as \cite{Apreda:2001us}
\begin{equation}
\beta \equiv -\frac{\mathrm{d}S_E}{\mathrm{d}t}\Bigg|_{t=t_n} \simeq \frac{1}{\Gamma}\frac{\mathrm{d}\Gamma}{\mathrm{d}t}\Bigg|_{t=t_n}.
\end{equation}
However, in the actual calculations, the renormalized parameter $\tilde{\beta}$ is more convenient:
\begin{equation}
\tilde{\beta} = T_n\frac{d}{dT}\left(\frac{S_3(T)}{T}\right)\Bigg|_{T=T_n}.
\end{equation}
The parameter $\alpha$ describes the strength of the phase transition, namely, the larger value of $\alpha$ corresponds to a stronger phase transition process.
In addition, the inverse of the parameter $\beta$ is related to the time scale of phase transition.
Based on the above approaches, we can numerically know the phase transition dynamics and calculate the phase transition parameters of all the benchmark point sets.

\section{Collider and Gravitational wave signatures}

After the three parameters $\alpha$, $\tilde{\beta} $ and $T_n$ are extracted from the finite-temperature effective potential,
we can predict the phase transition GW signals which are produced by three mechanisms:
bubbles collisions, sound waves, and magnetohydrodynamic turbulence in the plasma after collisions.
Based on the envelope approximation \cite{Kosowsky:1991ua,Kosowsky:1992rz,Kosowsky:1992vn,Kamionkowski:1993fg}, the numerical simulation gives the formula of the GW spectrum from bubble collisions \cite{Caprini:2015zlo,Huber:2008hg,Jinno:2016vai}:
\begin{equation}
h^2\Omega_{co}(f) \simeq 1.67\times10^{-5}\tilde{\beta}^{-2}\left(\frac{\kappa_{\phi}\alpha}{1 + \alpha}\right)^2\left(\frac{100}{g_{\star}}\right)^{1/3}\frac{0.11v_b^3}{0.42 + v_b^2}\frac{3.8(f/f_{co})^{2.8}}{1 + 2.8(f/f_{co})^{3.8}},
\end{equation}
where $g_{\star}$ is the total number of degrees of freedom at $T_n$, the coefficient $\kappa_{\phi}$ denotes the fraction of vacuum energy transformed into the gradient energy of the scaler fields, and $v_b$ is the bubble wall velocity.
The peak frequency is
\begin{equation}
f_{co} \simeq 1.65\times10^{-5}\text{Hz}\left(\frac{0.62}{1.8 - 0.1v_b + v_b^2}\right)\tilde{\beta}\left(\frac{T_n}{100GeV}\right)\left(\frac{g_{\star}}{100}\right)^{1/6}.
\end{equation}
The second source is generated by the sound waves of the bulk motion, and numerical simulation gives \cite{Hindmarsh:2013xza,Hindmarsh:2015qta}
\begin{equation}
h^2\Omega_{sw}(f) \simeq 2.65\times10^{-6}\tilde{\beta}^{-1}\left(\frac{\kappa_v\alpha}{1 + \alpha}\right)^2\left(\frac{100}{g_{\star}}\right)^{1/3}v_b(f/f_{sw})^3\left(\frac{7}{4 + 3(f/f_{sw})^2}\right)^{7/2},\label{swf}
\end{equation}
with the peak frequency
\begin{equation}
f_{sw} \simeq 1.9 \times10^{-5}\text{Hz}\frac{1}{v_b}\tilde{\beta}\left(\frac{T_n}{100GeV}\right)\left(\frac{g_{\star}}{100}\right)^{1/6},
\end{equation}
here $\kappa_v$ represents the fraction of vacuum energy that gets converted into bulk motion of the fluid.
The turbulence contribution to the GW spectrum is \cite{Caprini:2009yp,Binetruy:2012ze}
\begin{equation}
h^2\Omega_{turb}(f) \simeq 3.35\times10^{-4}\tilde{\beta}^{-1}\left(\frac{\kappa_{turb}\alpha}{1 + \alpha}\right)^{3/2}\left(\frac{100}{g_{\star}}\right)^{1/3}v_b\frac{(f/f_{turb})^3}{(1 + f/f_{turb})^{11/3}(1 + 8\pi f/h_{\star})},
\end{equation}
with the peak frequency
\begin{equation}
f_{turb} \simeq 2.7\times10^{-5}\text{Hz}\frac{1}{v_b}\tilde{\beta}\left(\frac{T_n}{100GeV}\right)\left(\frac{g_{\star}}{100}\right)^{1/6},
\end{equation}
and
\begin{equation}
h_{\star} = 1.65 \times 10^{-5}\text{Hz}\left(\frac{T_n}{100GeV}\right)\left(\frac{g_{\star}}{100}\right)^{1/6}.
\end{equation}
For the velocity profile of a Jouguet detonation front, the efficiency parameter $\kappa_v$, which is the ratio of bulk kinetic energy to the vacuum energy, is possible to be established as a function of $\alpha$ \cite{Espinosa:2010hh,Kamionkowski:1993fg}
\begin{equation}
\kappa_v = \frac{\sqrt{\alpha}}{0.135 + \sqrt{0.98 + \alpha}}\,\,.
\end{equation}
The precise calculation of bubble wall velocity is a very hard task, which involves microphysics and hydrodynamics. In the situation of bubble propagation by Jouguet detonation, the bubble wall velocity can be expressed as the following formula \cite{Espinosa:2010hh,Kamionkowski:1993fg}
\begin{equation}\label{vbn}
v_{b,J} = \frac{1/\sqrt{3} + (\alpha^2 + 2\alpha/3)^{1/2}}{1 + \alpha}\,\,\,,
\end{equation}
namely, $v_b = v_{b,J}$ for Jouguet detonation.
Reference~\cite{Espinosa:2010hh} shows that, as long as bubble propagates as detonation and deflagration mode, the gradient energy of the scaler  fields is negligible and most contribution to the GW signal is the bulk motion of the fluid. Hence, for Jouguet detonation and deflagration, $\kappa_{\phi}\rightarrow0$.

We also consider the deflagration mode with a bubble wall velocity $v_b = 0.5$ given by hand. In this case, we use the following efficiency parameter \cite{Espinosa:2010hh}
\begin{equation}
\kappa_v \approx \frac{c_s^{11/5}\kappa_A\kappa_B}{(c_s^{11/5} - v_b^{11/5})\kappa_B + v_bc_s^{6/5}\kappa_A},\label{effdf}
\end{equation}
where $c_s \cong 1/\sqrt{3}$ is the sound velocity and
\begin{equation}
\kappa_A \approx v_b^{6/5}\frac{6.9\alpha}{1.36 - 0.037\sqrt{\alpha} + \alpha},
\end{equation}
\begin{equation}
\kappa_B \approx \frac{\alpha^{2/5}}{0.017 + (0.997 + \alpha)^{2/5}}\,\,\,.
\end{equation}
	However, recent study \cite{Cutting:2019zws} shows the analytical fitting of the efficiency parameter for deflagration Eq.~\eqref{effdf} is not valid for all the parameter space of $v_b$ and $\alpha$, there can exist a large kinetic energy deficit in some parameter space.
	Therefore, the above modeling substantially overestimates the GW signals for deflagration.
	For both Jouguet detonation ($v_b > c_s$) and deflagration ($v_b < c_s$) mode, we set  $\kappa_{turb} \simeq 0.1\kappa_v$. However, the region where $v_b>c_s$ is not
all detonation. When $v_b$ is larger than $c_s$ but smaller than $v_{b,J}$ obtained by Eq.~\eqref{vbn}, it is called supersonic deflagration (hybrid) \cite{KurkiSuonio:1995pp, Espinosa:2010hh}. Here, we use $v_b > c_s$ to specify the Jouguet detonation mode. We do not study the supersonic deflagration mode in this work.
	Note we also use supersonic and subsonic to denote Jouguet detonation and deflagration in this section for simplicity.
We notice there is a new numerical simulation of the acoustic gravitational wave power spectrum \cite{Hindmarsh:2017gnf}, and gives some modifications to the formula of sound wave source used in this work.

Strong GW signal favors supersonic bubble wall velocity.
However, the EW baryogenesis prefers subsonic bubble wall velocity.
Actually, the bubble wall velocity obtained from Eq.~(\ref{vbn}) is not accurate enough here since these
formula is obtained in the simplest scalar model.
It is still possible that the real bubble wall velocity in this model is smaller than the velocity of sound wave for non-supercooling case.
To see the differences between the two choices of bubble wall velocity, we show the GW spectra of the same benchmark sets with a bubble wall velocity calculated by Eq.~(\ref{vbn}) and a fixed input subsonic velocity $v_b = 0.5$, respectively.

It is worthy noticing that the above formulas of the GW spectrum for the three sources, which are given by numerical simulation, are based on a rapid phase transition process and $\alpha < 1$. Since a supercooling FOPT may induce a longer and stronger transitions~\cite{Iso:2017uuu,Kobakhidze:2017mru}, it is not clear whether these formulas are applicable to this situation.
The GW spectrum induced by the supercooling phase transition is still controversial, we need a more detailed study.
Therefore, we just give the GW spectra of the benchmark sets without supercooling.

\begin{figure}[htp!!]
	\centering
	\subfigure{
		\begin{minipage}[t]{0.5\linewidth}
			\centering
			\includegraphics[scale=0.55]{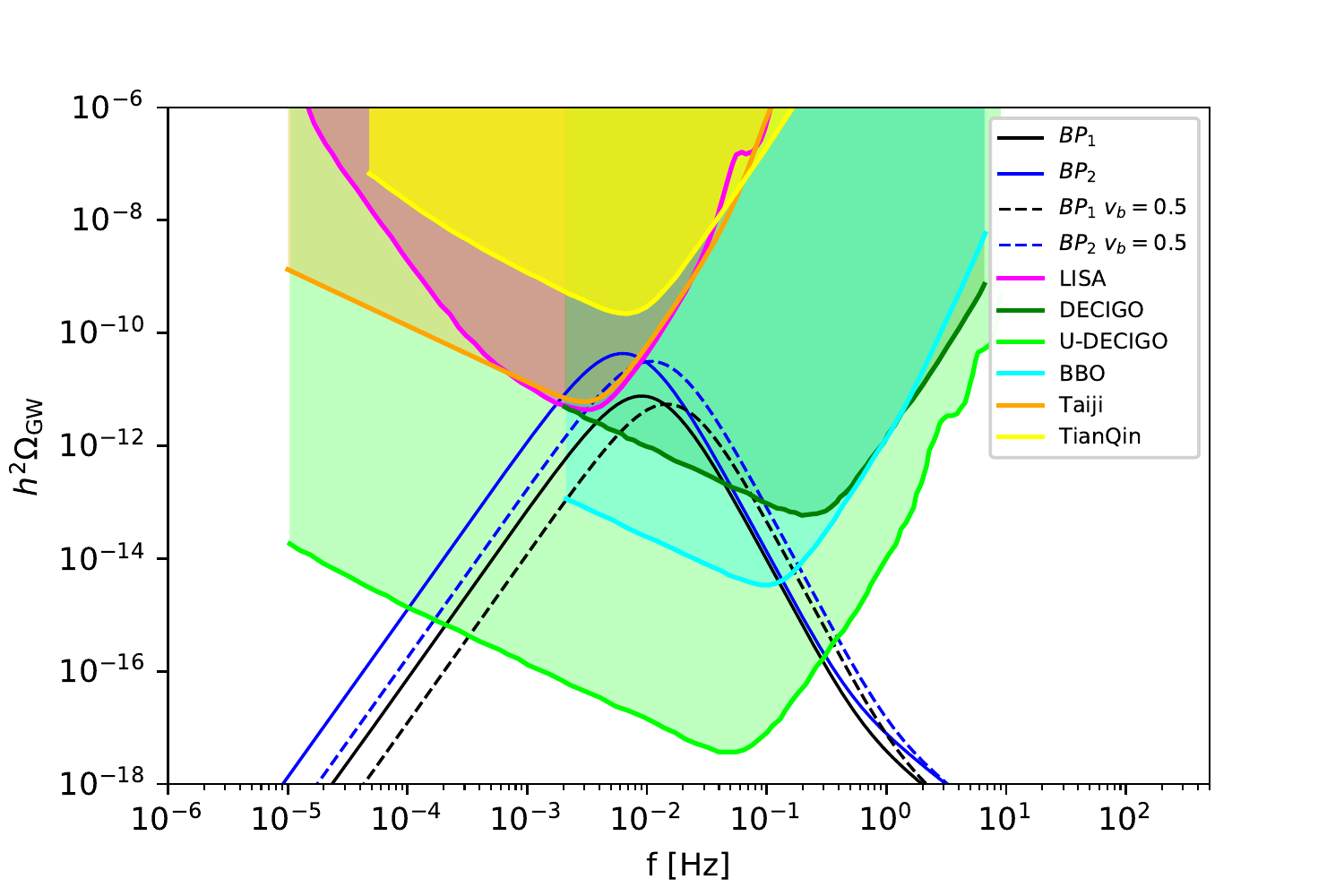}
	\end{minipage}}%
	\subfigure{
		\begin{minipage}[t]{0.5\linewidth}
			\centering
			\includegraphics[scale=0.55]{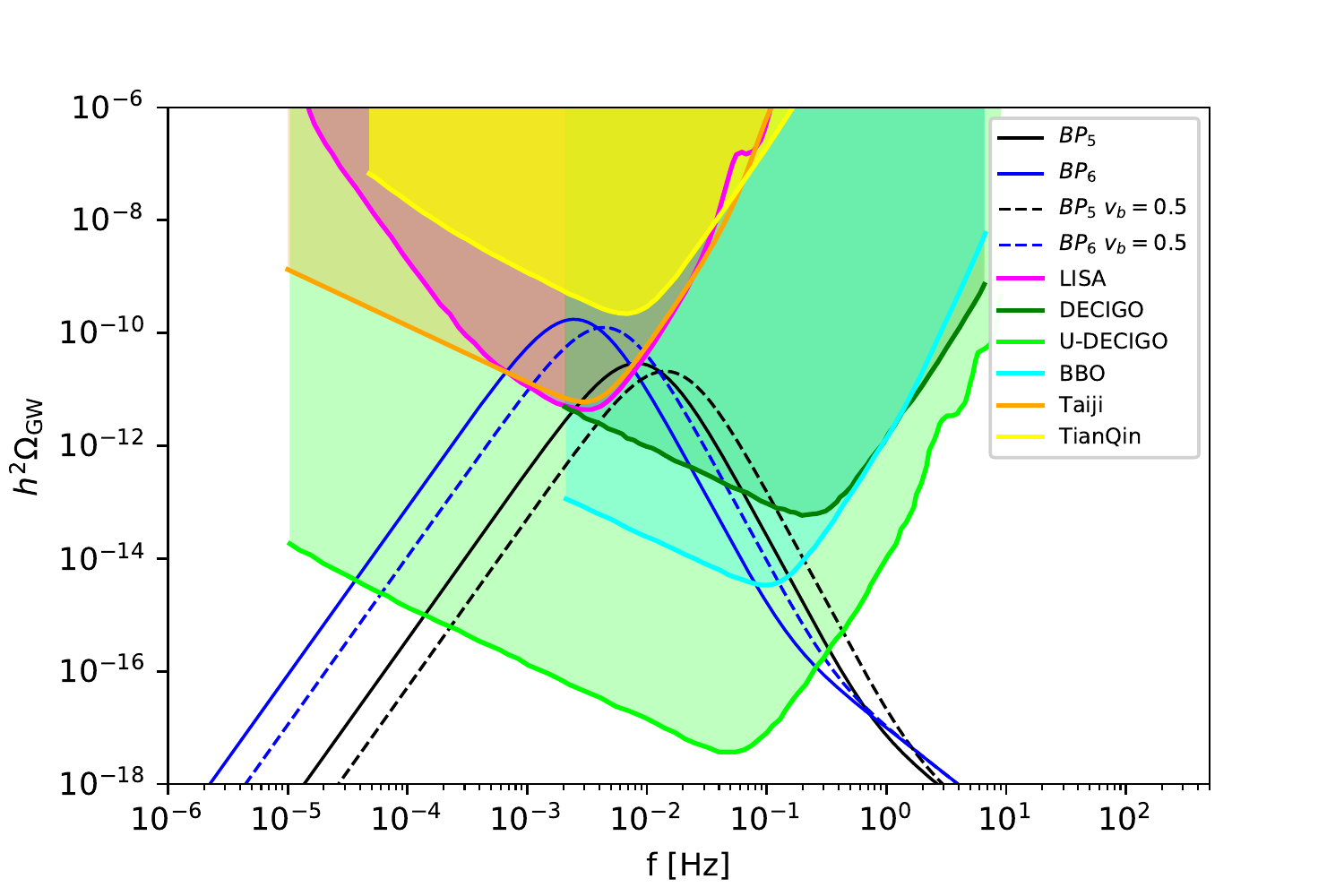}
	\end{minipage}}	
	\caption{
		The GW spectra of the one-step phase transition for the benchmark sets $BP_1$, $BP_2$, $BP_5$ and $BP_6$.
		The color shaded regions correspond to the expected sensitivity of the GW interferometers LISA, DECIGO, U-DECIGO, BBO, Taiji, and TianQin, respectively.
		The black and the blue solid lines denote the GW spectra of $BP_1$, $BP_2$, $BP_5$ and $BP_6$ with the bubble wall velocity determined by Eq.~(\ref{vbn}). The black and the blue dashed lines represent the GW spectra of $BP_1$, $BP_2$, $BP_5$ and $BP_6$ with a given subsonic bubble velocity $v_b= 0.5$.
		}\label{fg1}
\end{figure}

\begin{figure}[htp!!]
	\centering
	\subfigure{
		\begin{minipage}[t]{0.5\linewidth}
			\centering
			\includegraphics[scale=0.55]{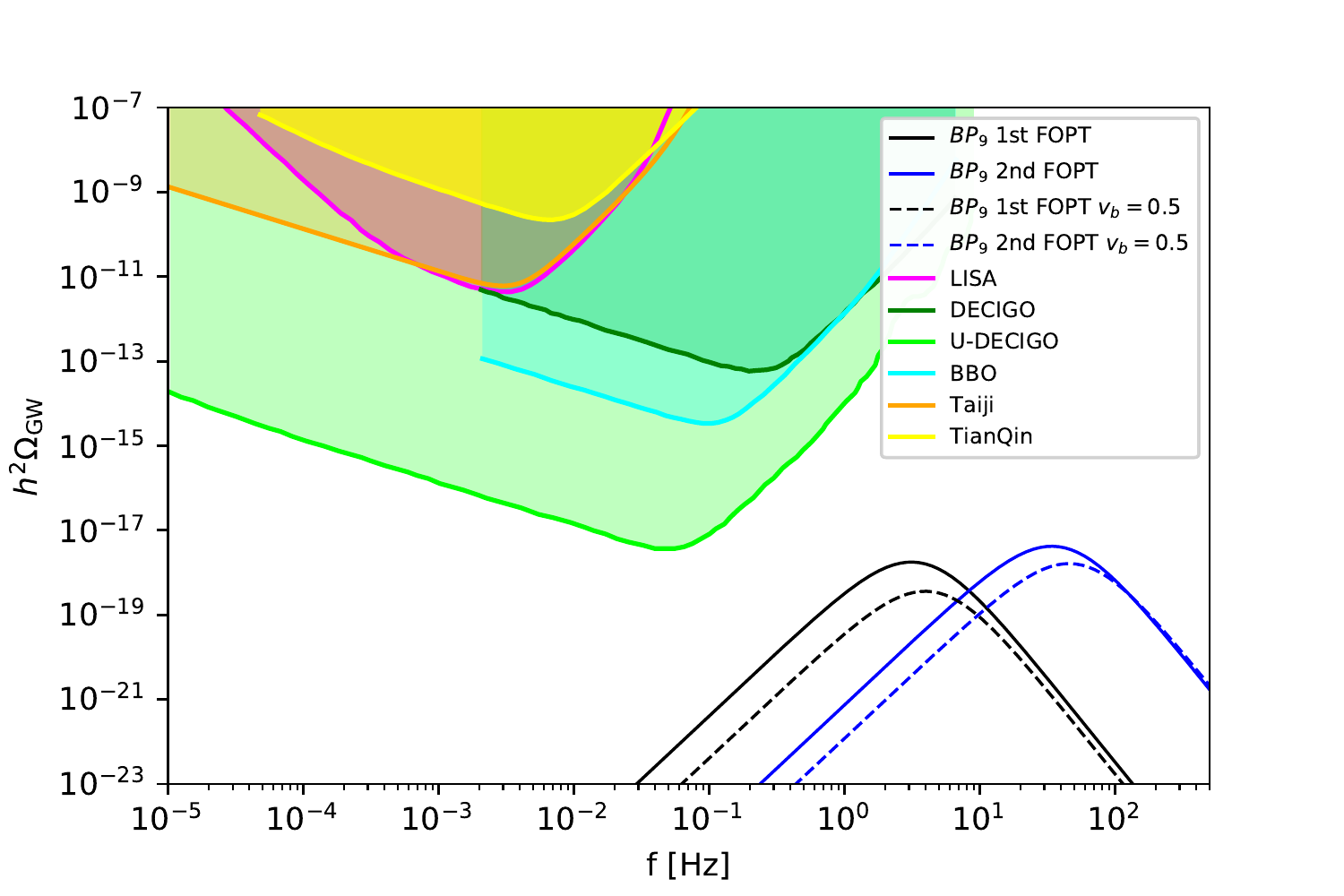}
	\end{minipage}}%
	\subfigure{
		\begin{minipage}[t]{0.5\linewidth}
			\centering
			\includegraphics[scale=0.55]{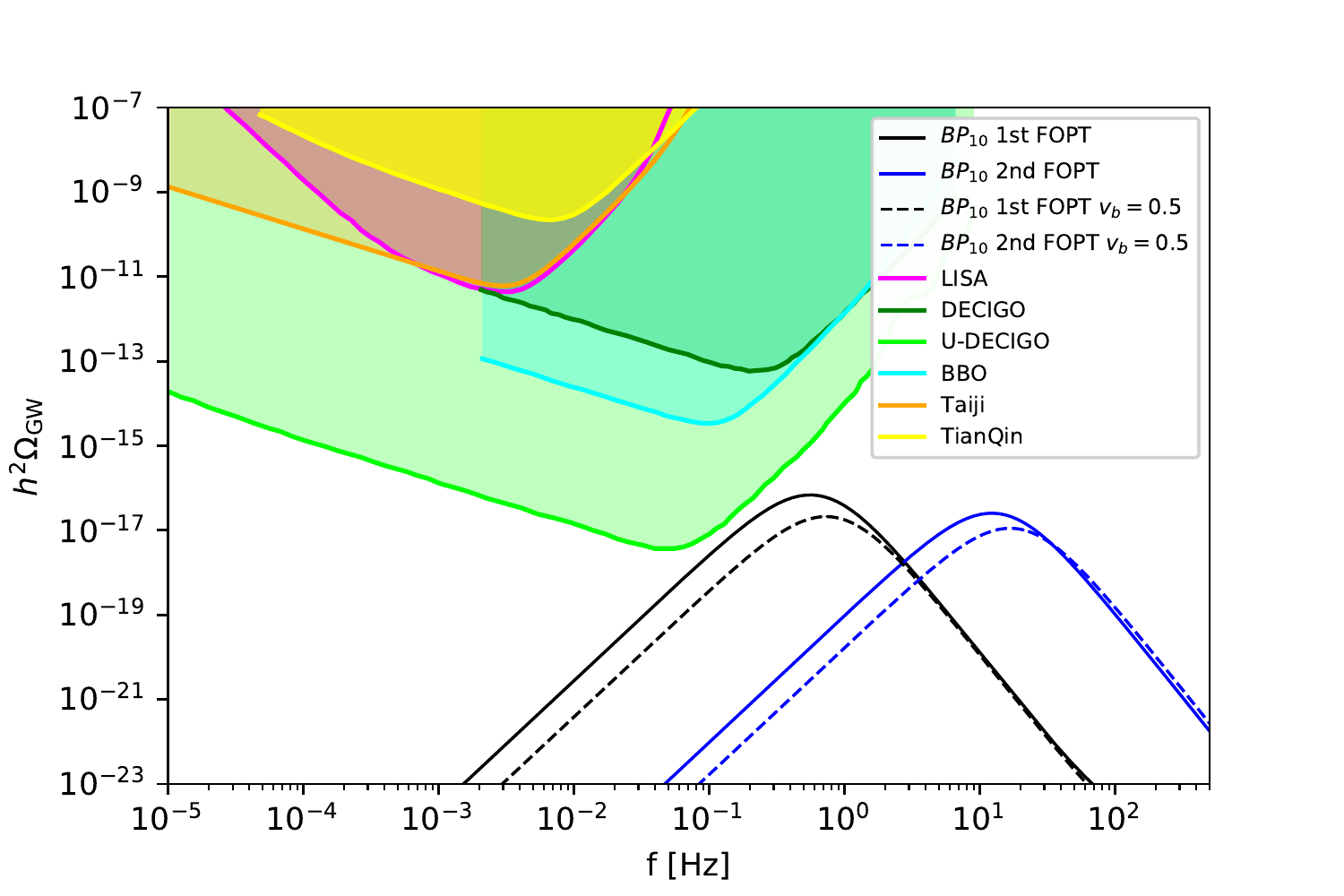}
	\end{minipage}}	
	\caption{
		The GW spectra of the two-step phase transition for $BP_9$ and $BP_{10}$.
		The black and the blue solid lines denote the GW spectra of the first and the second FOPT of $BP_9$ and $BP_{10}$ with bubble velocity calculated by Eq.~(\ref{vbn}). The black and the blue dashed lines denote the GW spectra of the first and the second FOPT of $BP_9$ and $BP_{10}$ with bubble velocity given by hand as $v_b = 0.5$.
		}\label{fg2}
\end{figure}

\begin{figure}[htp!!]
	\centering
	\subfigure{
		\begin{minipage}[t]{0.5\linewidth}
			\centering
			\includegraphics[scale=0.55]{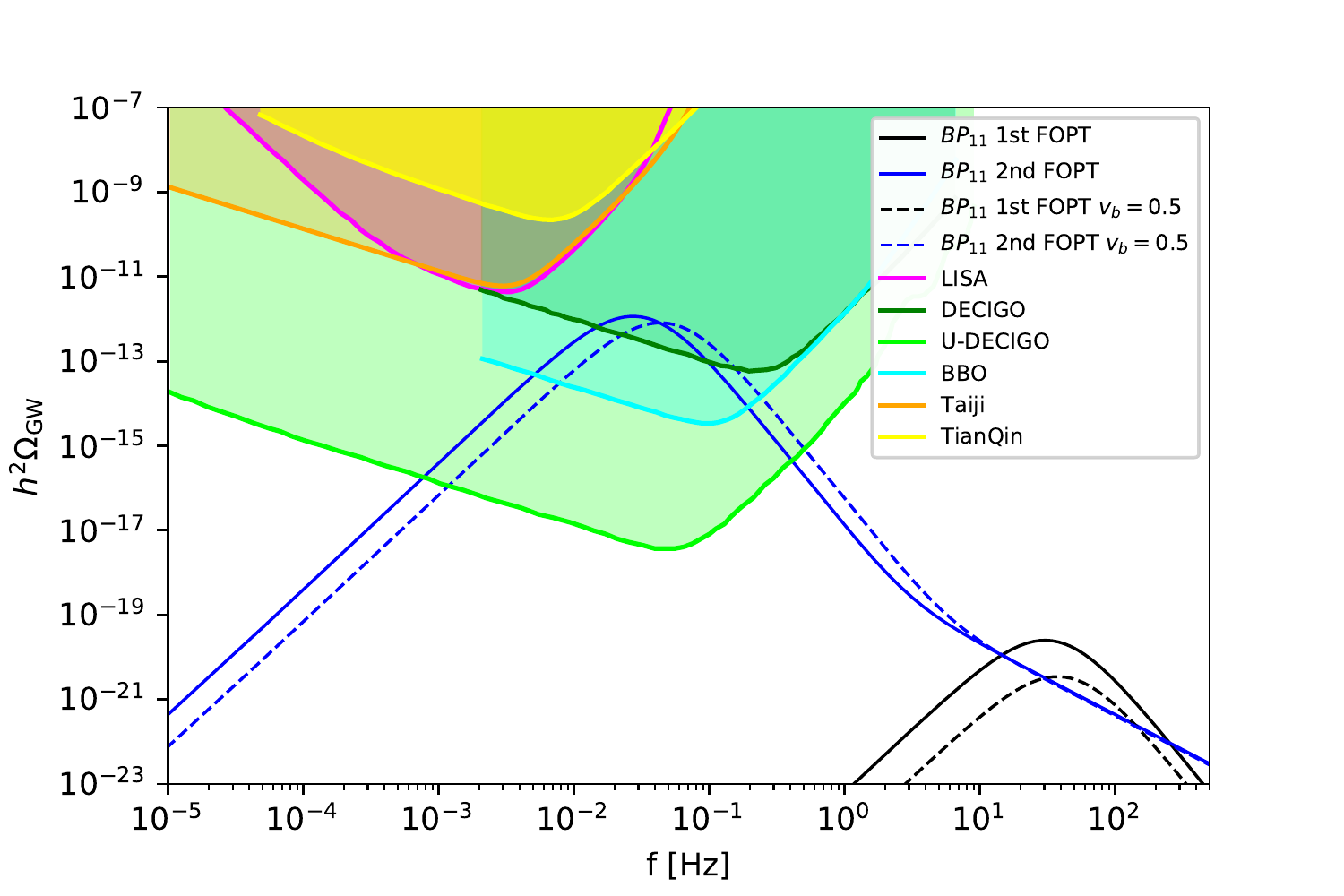}
	\end{minipage}}%
	\subfigure{
		\begin{minipage}[t]{0.5\linewidth}
			\centering
			\includegraphics[scale=0.55]{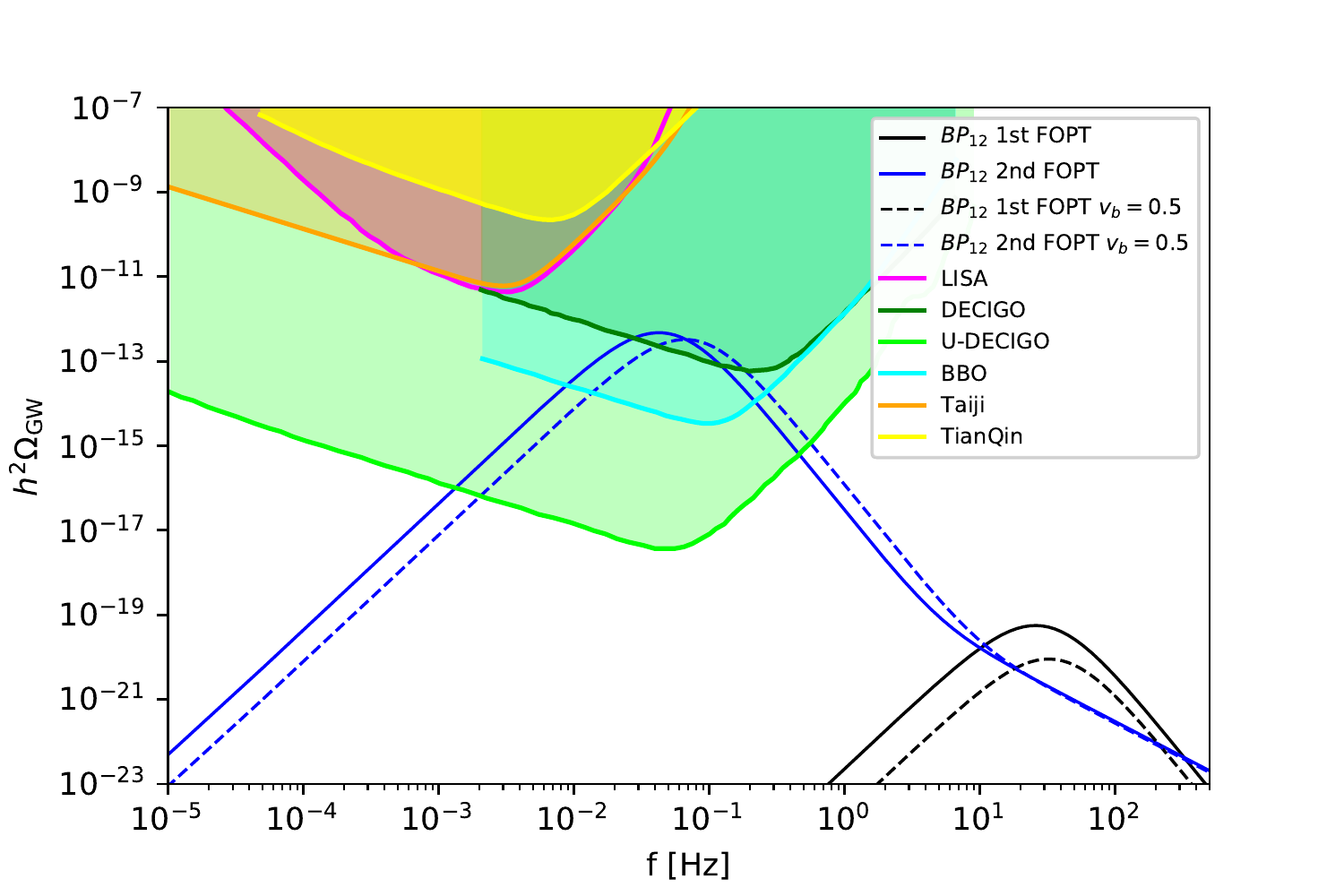}
	\end{minipage}}	
	\caption{
	The GW spectra of the three-step phase transition for $BP_{11}$ and $BP_{12}$.
	The black and the blue solid lines denote the GW spectra of the first and the second FOPT of $BP_{11}$ and $BP_{12}$ with bubble wall velocity calculated by Eq.~\eqref{vbn}.
	The black and the blue dashed lines denote the GW spectra of the first and the second FOPT of $BP_{11}$ and $BP_{12}$ with a bubble wall velocity $v_b = 0.5$.
	}
    \label{fg3}
\end{figure}

Since the energy in the scaler fields is negligible for detonation and deflagration, the GW signals mainly come from sound wave and turbulence.
Combining the two contributions, we show the numerical results of the approximated GW spectra in the C2HDM for the above benchmark sets in Fig.~\ref{fg1}, Fig.~\ref{fg2}, and Fig.~\ref{fg3}.
In each figure, we use the solid lines to represent the GW signals generated by the supersonic bubble wall velocity and the dashed lines to denote the signals induced by the subsonic bubble wall velocity.
For each benchmark set, the GW signal for the supersonic bubble wall velocity is stronger than
the GW signal for the subsonic case.
We also show the sensitivity curve of the planed GW interferometers LISA, DECIGO, U-DECIGO, BBO, Taiji, and TianQin in these figures.

Figure~\ref{fg1} shows the GW spectra of $BP_1$, $BP_2$, $BP_5$, and $BP_6$ that can induce a one-step FOPT.
Fig.~\ref{fg2} shows the GW spectra of $BP_9$ and $BP_{10}$ which can generate a two-step FOPT process.
Even though the signals are not strong enough compared to the proposed GW detector at current stage, they are still intriguing phase transition patterns.
They can produce two copies of GW signals with different peak frequencies~\cite{Wan:2018udw,Addazi:2017nmg,Huang:2017laj,Athron:2019teq}.
Their signals are different from the one-step FOPT as shown in Fig.~\ref{fg1}, where there exists only one copy of GW signals for given benchmark sets.
In Fig.~\ref{fg3}, the GW spectra of the benchmark points with three-step phase transition.
The GW signal from the first FOPT of the three-step phase transition is much more weaker than the second one.

\begin{table}[htp]
	\centering
	\begin{tabular}{ccccccccc}
		\hline\hline
		& \multicolumn{2}{c}{$BP_1$}& \multicolumn{2}{c}{$BP_2$} & \multicolumn{2}{c}{$BP_5$}& \multicolumn{2}{c}{$BP_6$}\\
		& $v_b>c_s$ & $v_b<c_s$ & $v_b>c_s$ & $v_b<c_s$ & $v_b>c_s$ & $v_b<c_s$ & $v_b>c_s$ & $v_b<c_s$\\
		\hline
		$SNR_{(LISA)}$& 14.17 & 4.05 & 131.15 & 42.39 & 61.94 & 16.64 & 716.49 & 501.38 \\
		$SNR_{(Taiji)}$& 12.54 & 3.60 & 118.87 & 37.43 & 55.00 & 14.76 & 765.20 & 480.38 \\
		$SNR_{(TianQin)}$& 1.62 & 0.74 & 10.46 & 5.83 & 6.47 & 2.95 & 28.82 & 28.98 \\
		\hline\hline
	\end{tabular}
	\caption{The SNR  of the GW signals in C2HDM for the one-step phase transition benchmark sets which could be detected by the planed LISA, Taiji, and TianQin experiments. $v_b > c_s$ and $v_b<c_s$ represent Jouguet detonation and deflagration case, respectively.}\label{SNR}
\end{table}

To claim a detection of a GW signal, the quantity signal-to-noise ratio (SNR) is defined,
\begin{equation}\label{snr}
\text{SNR}=\sqrt{\mathcal{T} \int^{f_{max}}_{f_{min}} df \left( \frac{h^2 \Omega_{GW}}{h^2  \Omega_{sens}}\right)^2}\,\,,  \nonumber
\end{equation}
where $h^2\Omega_{sens}$ represents the sensitivity of a given experiment configuration and
$\mathcal{T}$ is the duration time of the mission. The signal is detectable if SNR is larger than
a threshold $\rm SNR_{thre}$, which is not a easy task to be quantified. Based on Ref.~\cite{Caprini:2015zlo}, we use $\rm SNR_{thre} =10$.
According to \cite{Caprini:2019egz,LISA:documents}, for $\mathcal{T}$,
we choose 4 years as the mission duration and a duty cycle of $75\%$,
yielding $\mathcal{T}\simeq 9.46\times10^7s$, which is the minimal data-taking time guaranteed by
LISA.
We quantify the reaches of the planed experiments (LISA, Taiji, and TianQin) by calculating the SNR of the one-step phase transition benchmark sets, as shown in Table~\ref{SNR}.
	
Besides the detectable GW signals, the strong FOPT could also induce obvious deviation of Higgs trilinear coupling compared to the SM~\cite{Kanemura:2004ch}, which can be parametrized as the following form
\begin{equation}
\mathcal{L}_{HHH}=-\frac{1}{3!} (1+\delta_H ) A_H H^3 \,\,,
\end{equation}
where $A_H$ is the Higgs trilinear coupling in the SM.
Here and below, we use $H$ to label the Higgs boson instead of $H_1$ for simplicity.
In Table~\ref{fp1}, Table~\ref{fp2} and Table~\ref{fp3}, we show the deviation of the Higgs trilinear coupling for each benchmark set.
The deviation of Higgs trilinear coupling $\delta_H$ from SM roughly varies from 1.049 to 1.863 at one-loop level for these benchmark points, which are calculated using the package \emph{BSMPT}~\cite{Basler:2017uxn,Basler:2018cwe}.
At LHC, it may be not easy to pin down this deviation from Higgs pair production~\cite{Bian:2016awe} due to the large SM backgrounds.
High luminosity LHC may improve the sensitivity~\cite{Chala:2018opy}.
However, the significant modification of Higgs trilinear coupling may be measured by the Higgs boson pair production at future high energy hadron collider.
This obvious deviation can also modify the cross section of $e^+e^- \to Z H$ process through loop contributions.
Therefore, it can be indirectly tested by the precise measurements of the cross section for the $Z$ boson and Higgs boson associated production at the future lepton collider, such as CEPC or ILC and FCC-ee~\cite{Huang:2015izx,Huang:2016odd,Huang:2017rzf,Cao:2017oez,Huang:2018aja}.
The deviation of the $ZH$ cross section can be defined as
\begin{equation}
\delta(ZH)=\frac{\sigma_{ZH}^{\rm C2HDM}}{\sigma_{ZH}^{\rm SM}}-1
\end{equation}
At 240 GeV CEPC with 5.6 $ab^{-1}$ integrated luminosity, the estimated precision of $\sigma_{ZH}$ is about $0.5\%$,
which means all the benchmark sets are within the sensitivity of
CEPC~\cite{CEPCStudyGroup:2018ghi}. The sensitivity for FCC-ee is even better, about $0.4\%$.
The corresponding numerical results for each benchmark set are shown in Table~\ref{fp1}, Table~\ref{fp2} and Table~\ref{fp3}.
In the tables, each benchmark set corresponds to $\alpha, \tilde{\beta}, T_n$ (they determine the GW signal) and $\delta_H$ (it determines
the collider signal), which means the GW signal and collider signal are correlated by same set of parameter of the EW phase transition dynamics.
Therefore, the future lepton colliders in complementary to GW experiments~\cite{Huang:2015izx,deVries:2017ncy,Huang:2016odd,Huang:2017rzf,Cao:2017oez,Huang:2018aja} can help to unravel different
phase transition dynamics. Namely, these two complementary experiments can help us to understand whether the phase transition
process is one-step FOPT, or two-step FOPTs or even three-step phase transitions in the early universe.
\begin{table}[htp!!]
	\centering
	\begin{tabular}{ccccccccc}
		\hline\hline
		&pattern& $T_n$ [GeV]& $\epsilon(T_n)[\rm GeV^4]$ & $v_b$ & $\alpha$ & $\tilde{\beta}$ &$\delta_H@$one-loop & $\delta(ZH)$\\
		\hline
		$BP_1$ & 1-step & 59.653 & $6.892\times10^{7}$ & 0.825 & 0.192 & 648.048 & 1.135 &1.816\% \\
		$BP_2$ & 1-step & 45.291 & $4.493\times10^{7}$ & 0.875 & 0.376 & 630.773 & 1.338 &2.141\%  \\
		$BP_3$ & 1-step & 25.964 & $2.771\times10^{7}$ & 0.964 & 2.149 & 471.699 &   1.677 & 2.684\% \\
		$BP_4$ & 1-step & 23.644 & $2.714\times10^{7}$ & 0.974 & 3.060 & 414.956 &   1.723   &2.737\% \\
        $BP_5$ & 1-step & 40.912 & $2.954\times10^{7}$ & 0.874 & 0.372 & 915.233   & 1.652  & 2.643\% \\
		$BP_6$ & 1-step & 36.639 & $2.61\times10^{7}$ & 0.895 & 0.510 & 313.287   & 1.672  & 2.674\% \\
		$BP_7$ & 1-step & 26.529 & $2.121\times10^7$ & 0.952 & 1.509 & 100.331 & 1.720  & 2.752\% \\
		$BP_8$ & 1-step & 27.621 & $2.188\times10^7$ & 0.947 & 1.325 & 81.825 & 1.680 & 2.687\% \\
		\hline\hline
	\end{tabular}
	%	\captionsetup{type=table}
	\caption{Correlation between the GW parameters ($\alpha, \tilde{\beta}, T_n$) and the collider parameter (the modification of Higgs trilinear coupling at one loop $\delta_H$) for the one-step phase transition pattern. $\delta(ZH)$ represents the corresponding loop-induced modification of ZH cross section at 240 GeV CEPC.}\label{fp1}
\end{table}

\begin{table}[htp!!]
	\centering
	\begin{tabular}{ccccccccc}
		\hline\hline
		& pattern & $T_n$ [GeV]& $\epsilon(T_n)[\rm GeV^4]$ & $v_b$ & $\alpha$ & $\tilde{\beta}$   &$\delta_H@$one-loop & $\delta(ZH)$\\
		\hline
		\multirow{2}*{$BP_9$}
		& \multirow{2}*{2-step} & 96.995 & $1.532\times10^{7}$  & 0.638 & 0.00610 & 107292.81  & \multirow{2}*{1.049} & \multirow{2}*{1.678\%}  \\
		&  & 93.997 & $4.077\times10^{7}$ & 0.677 & 0.0184 & 1279659.55   &  &  \\ \hline
		\multirow{2}*{$BP_{10}$}
		& \multirow{2}*{2-step} & 93.462 & $2.56\times10^7$ & 0.659 & 0.0118 & 20542.25 & \multirow{2}*{1.104} & \multirow{2}*{1.766\%} \\
		&  & 91.920 & $4.892\times10^7$ & 0.690 & 0.0241 & 479401.89 &  &  \\
		\hline\hline	
	\end{tabular}
	%	\captionsetup{type=table}
	\caption{Correlation between the GW parameters ($\alpha, \tilde{\beta}, T_n$) and the collider parameter (the modification of Higgs trilinear coupling at one loop $\delta_H$) for the two-step phase transition pattern (two consecutive FOPTs at different temperature). $\delta(ZH)$ represents the corresponding loop-induced modification of ZH cross section at 240 GeV CEPC.}\label{fp2}
\end{table}

\begin{table}[htp!!]
	\centering
	\begin{tabular}{ccccccccc}
		\hline\hline
		&  pattern & $T_n$ [GeV]& $\epsilon(T_n)[\rm GeV^4]$ & $v_b$ & $\alpha$ & $\tilde{\beta}$   &$\delta_H@$one-loop & $\delta(ZH)$\\
		\hline
		\multirow{2}*{$BP_{11}$}
		& \multirow{2}*{3-step} & 68.046 & $2.15\times10^{6}$  & 0.624 & 0.00353 & 1457261.58  & \multirow{2}*{1.863} & \multirow{2}*{2.980\%}  \\
		&  & 51.316 & $2.966\times10^{7}$ & 0.807 & 0.151 & 2235.16   &  &  \\ \hline
		\multirow{2}*{$BP_{12}$}
		& \multirow{2}*{3-step} & 69.380 & $2.864\times10^6$ & 0.629 & 0.00436 & 1225417.53 & \multirow{2}*{1.854} & \multirow{2}*{2.966\%} \\
		&  & 55.586 & $3.354\times10^7$ & 0.792 & 0.124 & 3142.96 &  &  \\
		\hline\hline	
	\end{tabular}
	%	\captionsetup{type=table}
	\caption{Correlation between the GW parameters ($\alpha, \tilde{\beta}, T_n$) and the collider parameter (the modification of Higgs trilinear coupling at one loop $\delta_H$) for the three-step phase transition pattern (two FOPTs and one SOPT).  $\delta(ZH)$ represents the corresponding loop-induced modification of ZH cross section at 240 GeV CEPC.}\label{fp3}
\end{table}

\section{Discussions}
For present study, we have not checked whether the CP-violating source is enough for EW baryogenesis and whether the sphaleron process is sufficiently quenched (We can see $\tilde{v}_1/T_n$ and $\tilde{v}_2/T_n$ from the evolution of $\tilde{v}_1$ an $\tilde{v}_2$ with the temperature in Fig.~\ref{wx1}, Fig.~\ref{wx3}, and Fig.~\ref{wx4}). Especially, for multi-step phase transition case, they are no necessarily responsible for EW baryogenesis. In this work, we focus on the phase transition dynamics, the corresponding GW signals in synergy with new collider signals at future lepton colliders. We leave the study of realization of EW baryogenesis in our future work. And we give brief discussion on the assumption of the dynamical CP-violation and supercooling.
\subsection{Consistent check on our assumptions: The evolution of the dynamical CP-violation}
\begin{figure}
	\centering
	\subfigure{
		\begin{minipage}[t]{0.5\linewidth}
			\centering
			\includegraphics[scale=0.55]{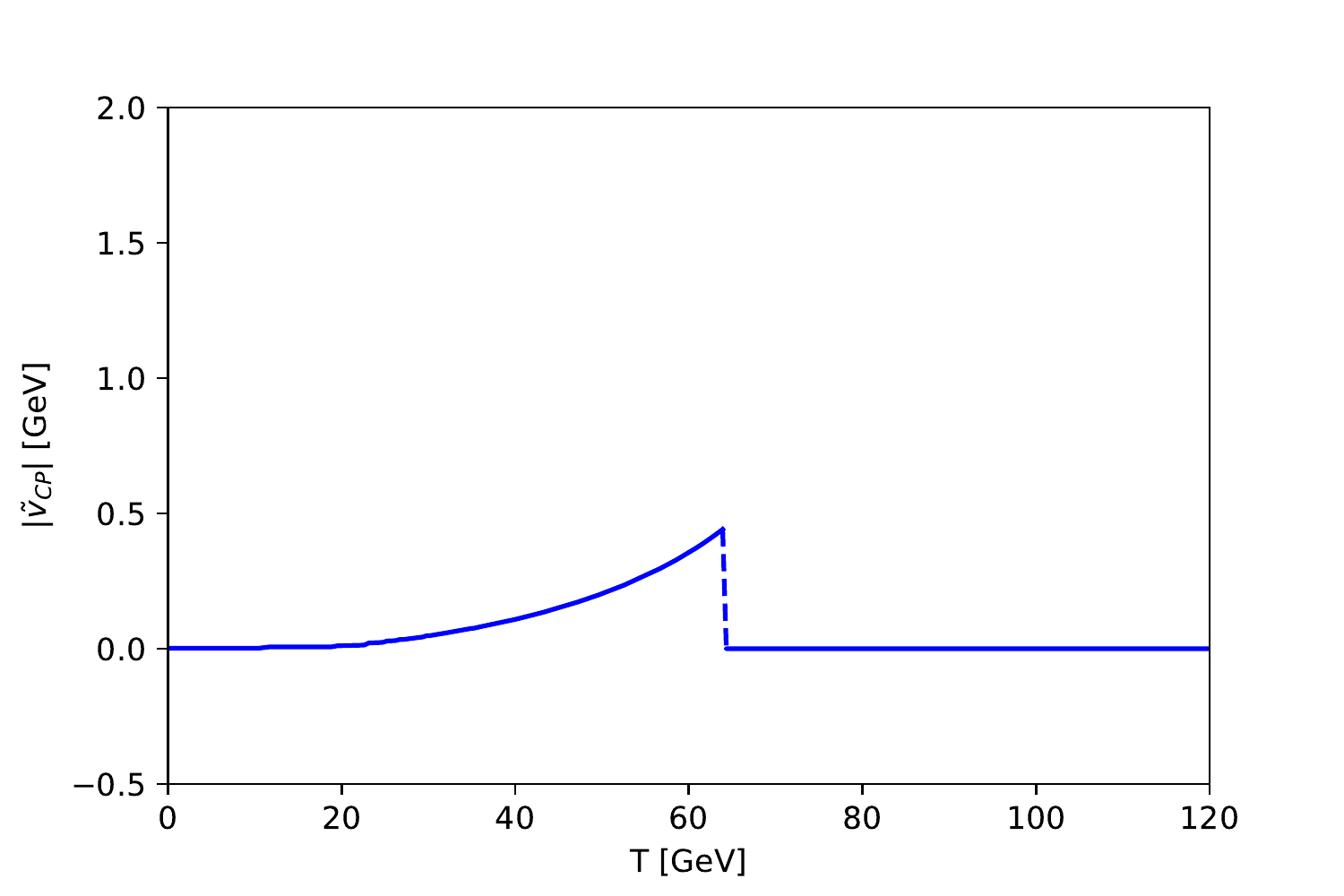}
	\end{minipage}}%
	\subfigure{
		\begin{minipage}[t]{0.5\linewidth}
			\centering
			\includegraphics[scale=0.55]{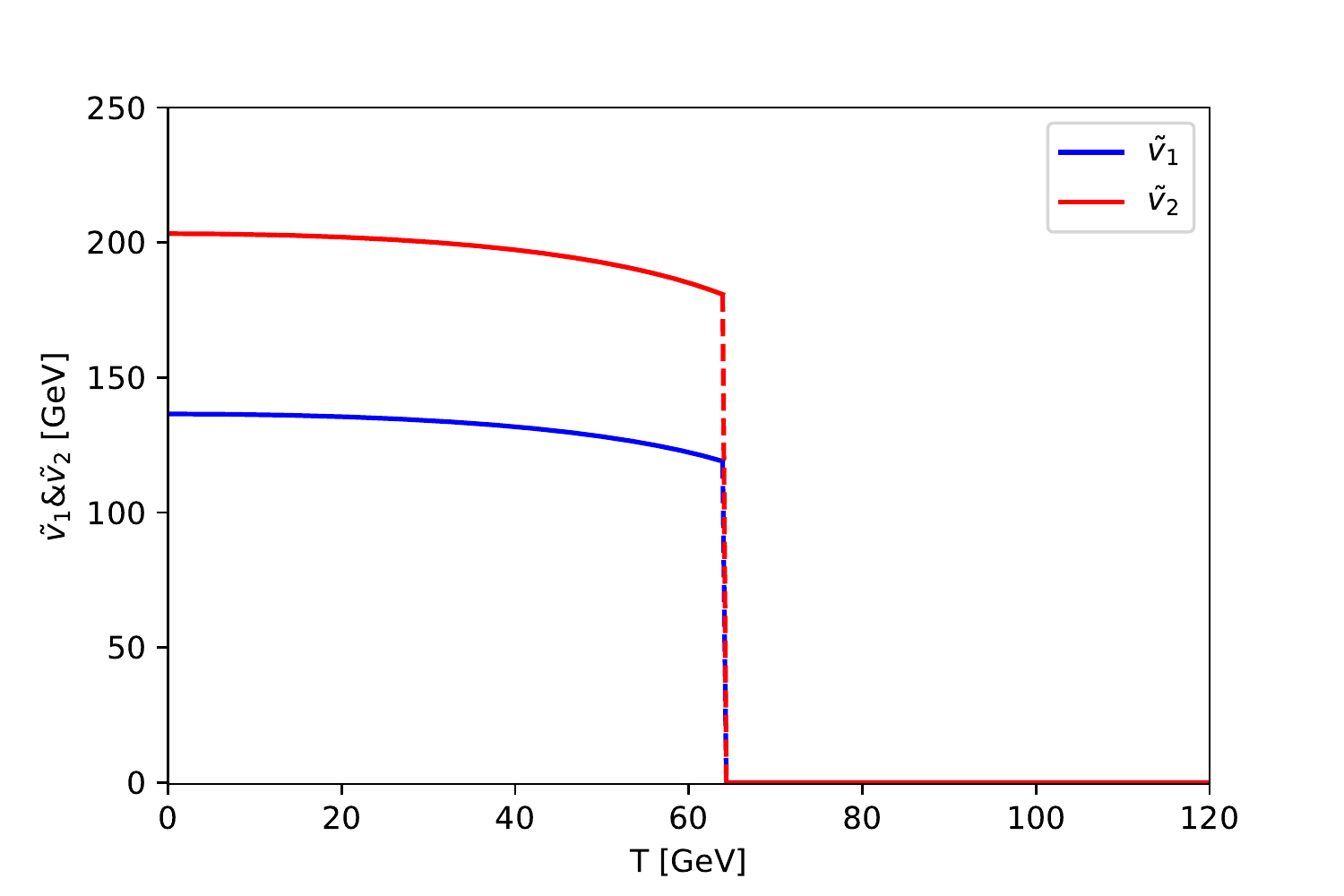}
	\end{minipage}}	
	\caption{Evolution of $\tilde{v}_{CP}$, $\tilde{v}_1$, and $\tilde{v}_2$ for one-step phase transition.}\label{wx1}
\end{figure}

\begin{figure}
	\centering
	\subfigure{
		\begin{minipage}[t]{0.5\linewidth}
			\centering
			\includegraphics[scale=0.55]{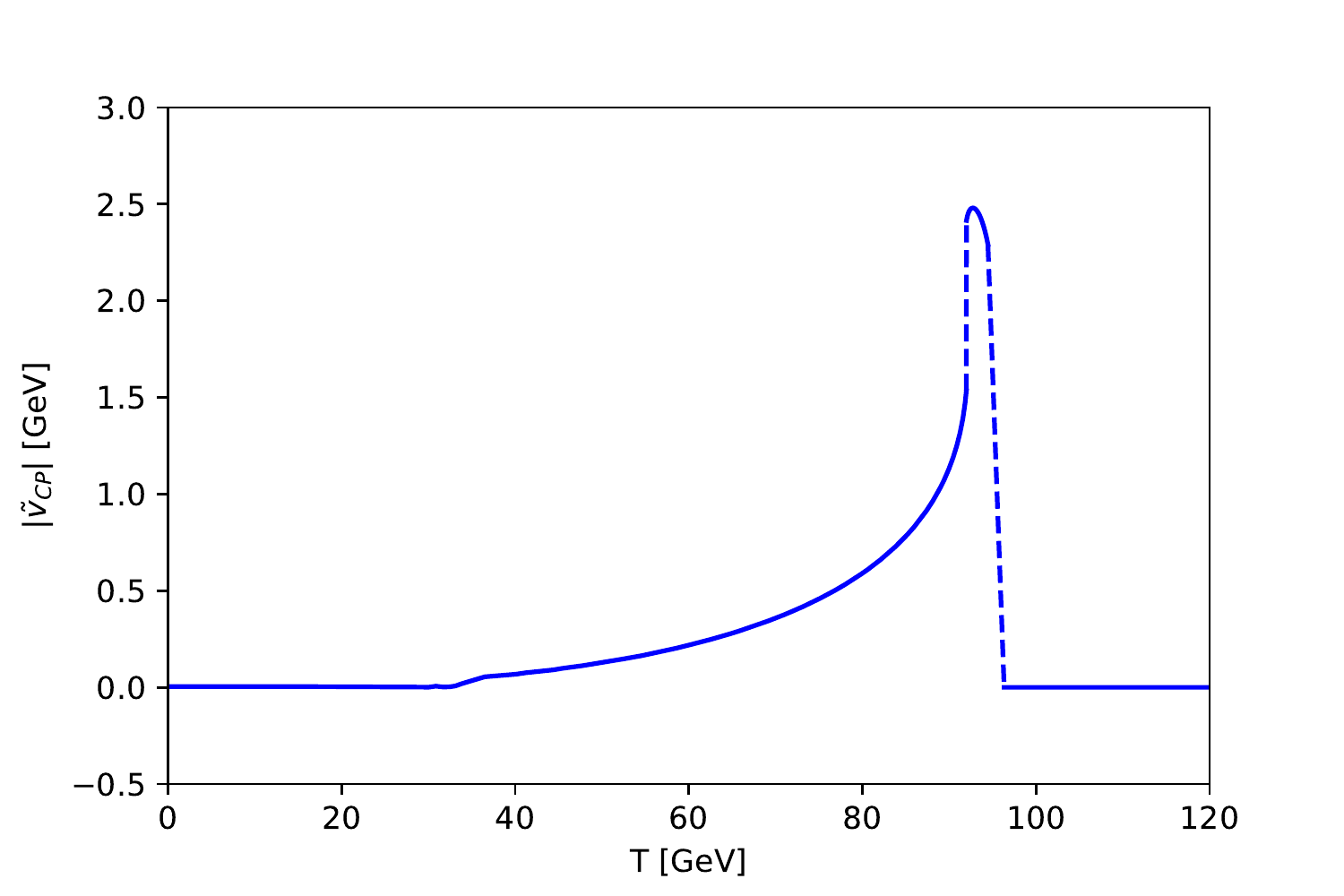}
	\end{minipage}}%
	\subfigure{
		\begin{minipage}[t]{0.5\linewidth}
			\centering
			\includegraphics[scale=0.55]{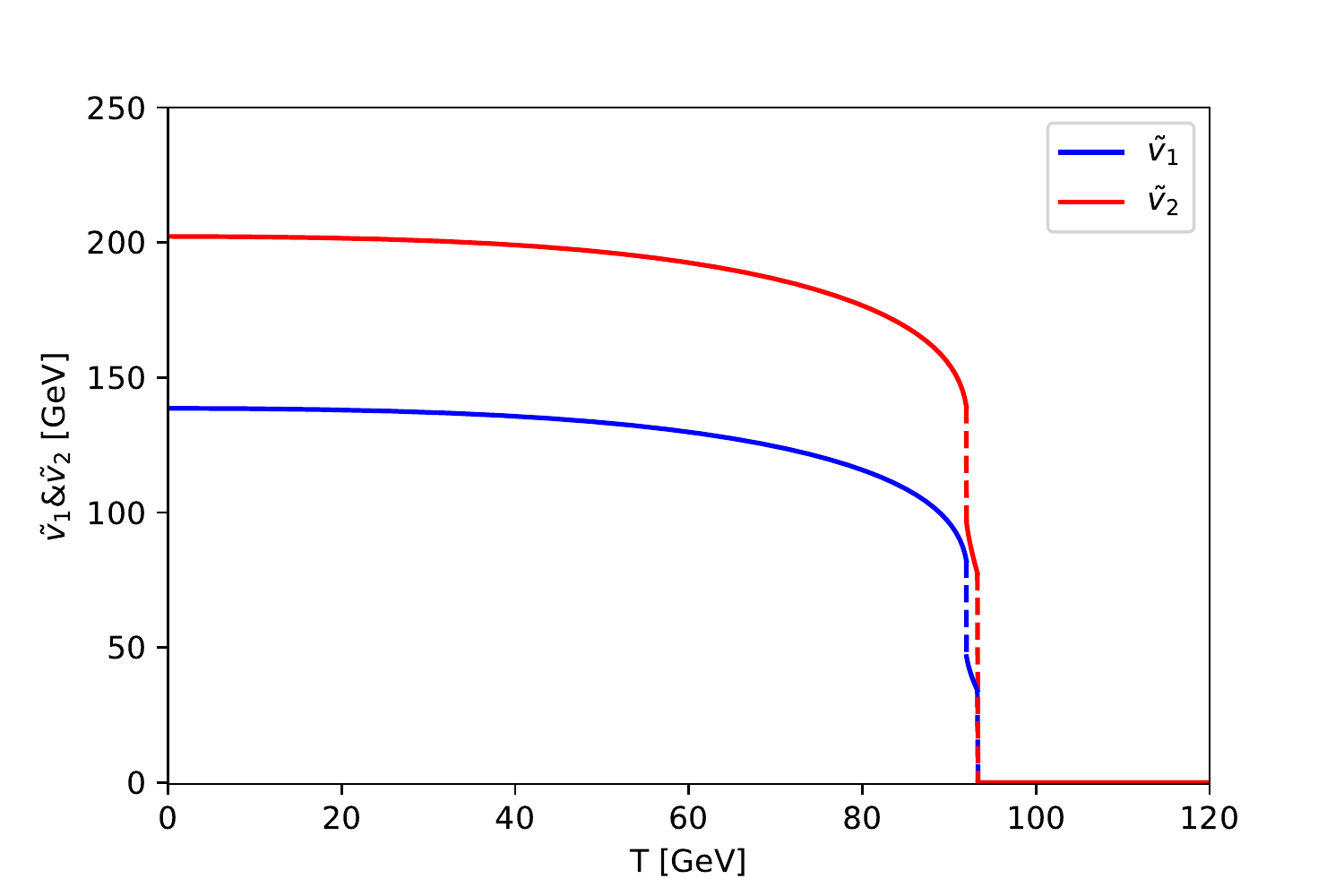}
	\end{minipage}}	
	\caption{Evolution of $\tilde{v}_{CP}$, $\tilde{v}_1$, and $\tilde{v}_2$ for two-step phase transition with two FOPTs.}\label{wx3}
\end{figure}

\begin{figure}
	\centering
	\subfigure{
		\begin{minipage}[t]{0.5\linewidth}
			\centering
			\includegraphics[scale=0.55]{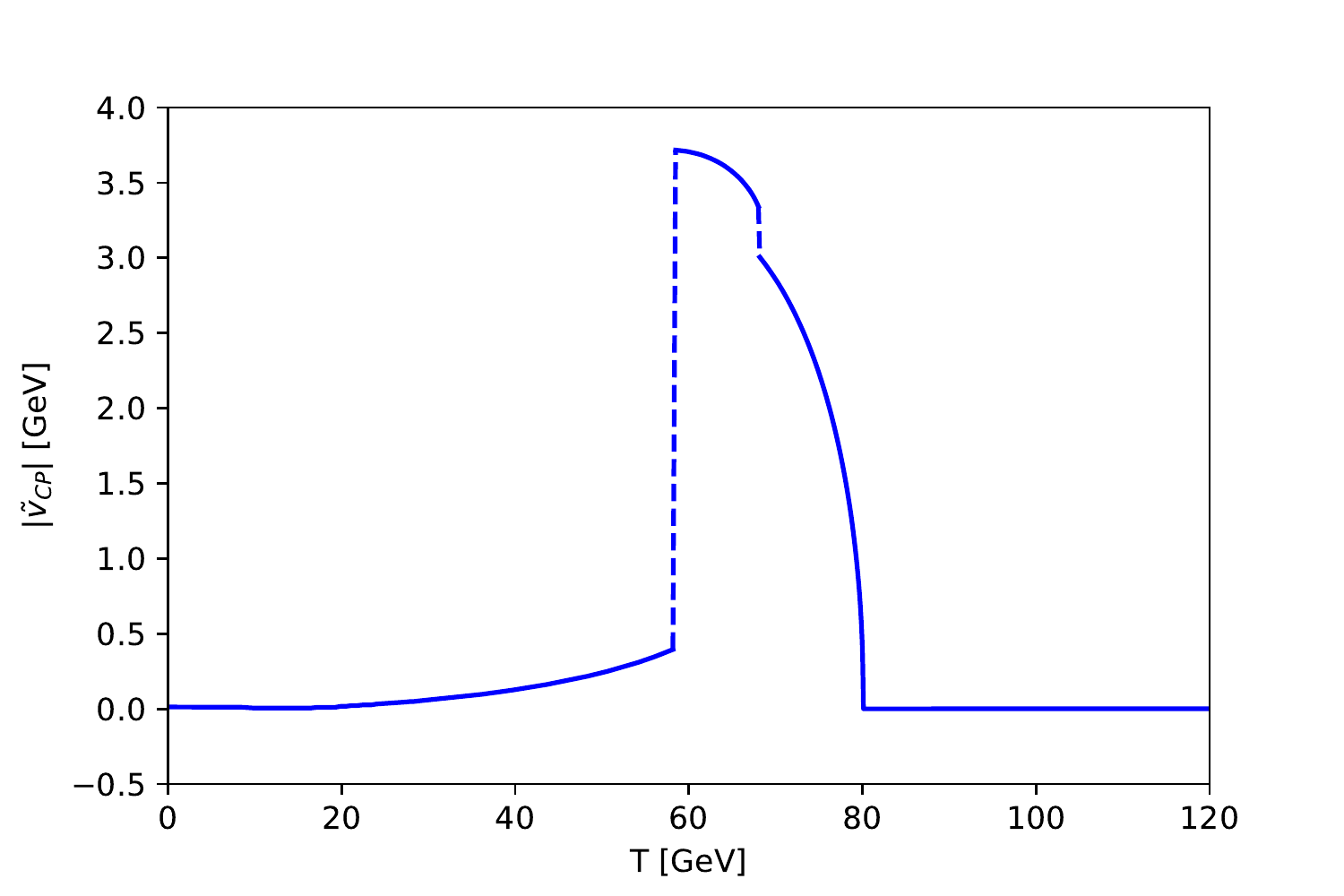}
	\end{minipage}}%
	\subfigure{
		\begin{minipage}[t]{0.5\linewidth}
			\centering
			\includegraphics[scale=0.55]{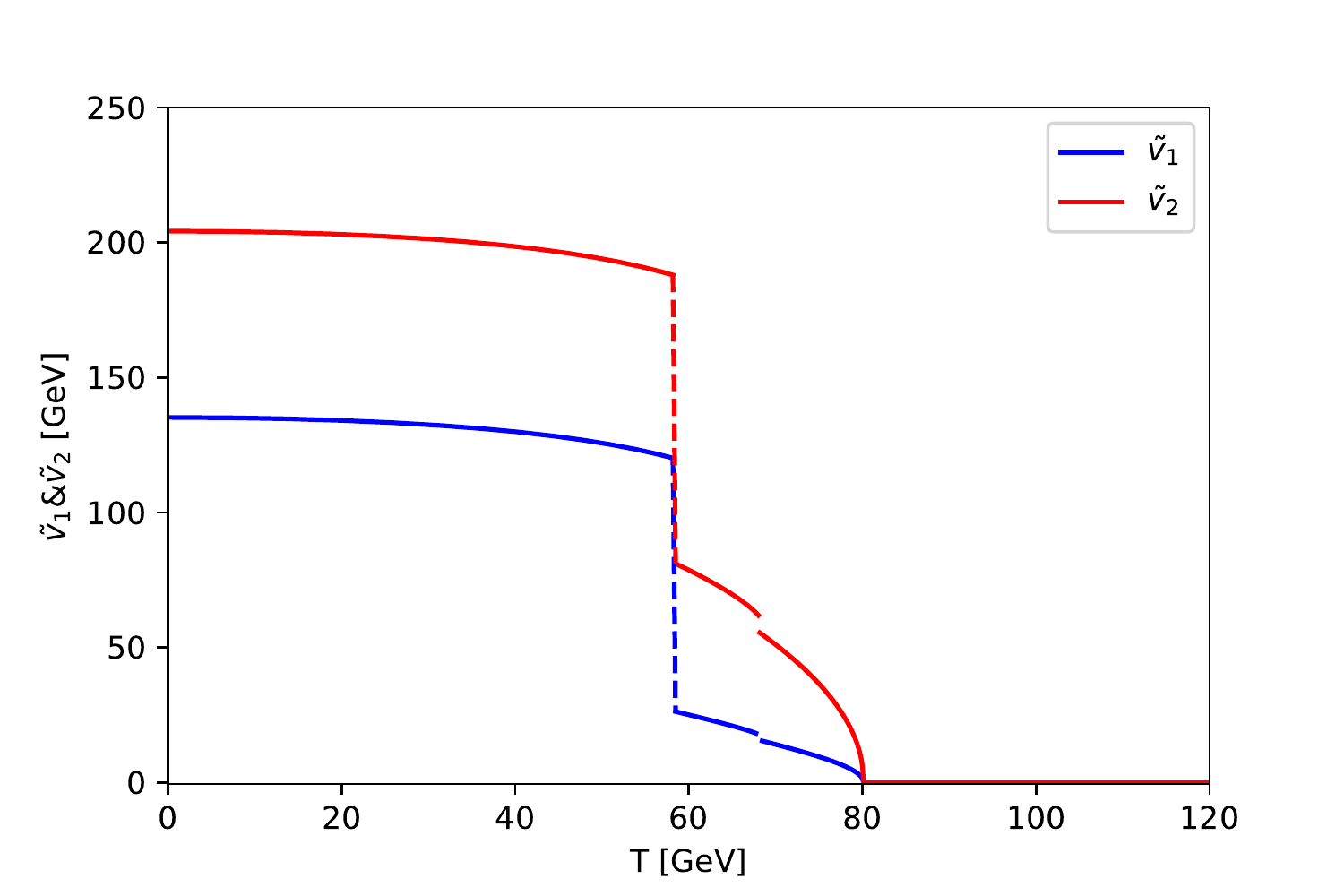}
	\end{minipage}}	
	\caption{Evolution of $\tilde{v}_{CP}$, $\tilde{v}_1$, and $\tilde{v}_2$ for three-step phase transition with two FOPTs and one SOPT.}\label{wx4}
\end{figure}

\begin{figure}
	\centering
	\includegraphics[scale=0.8]{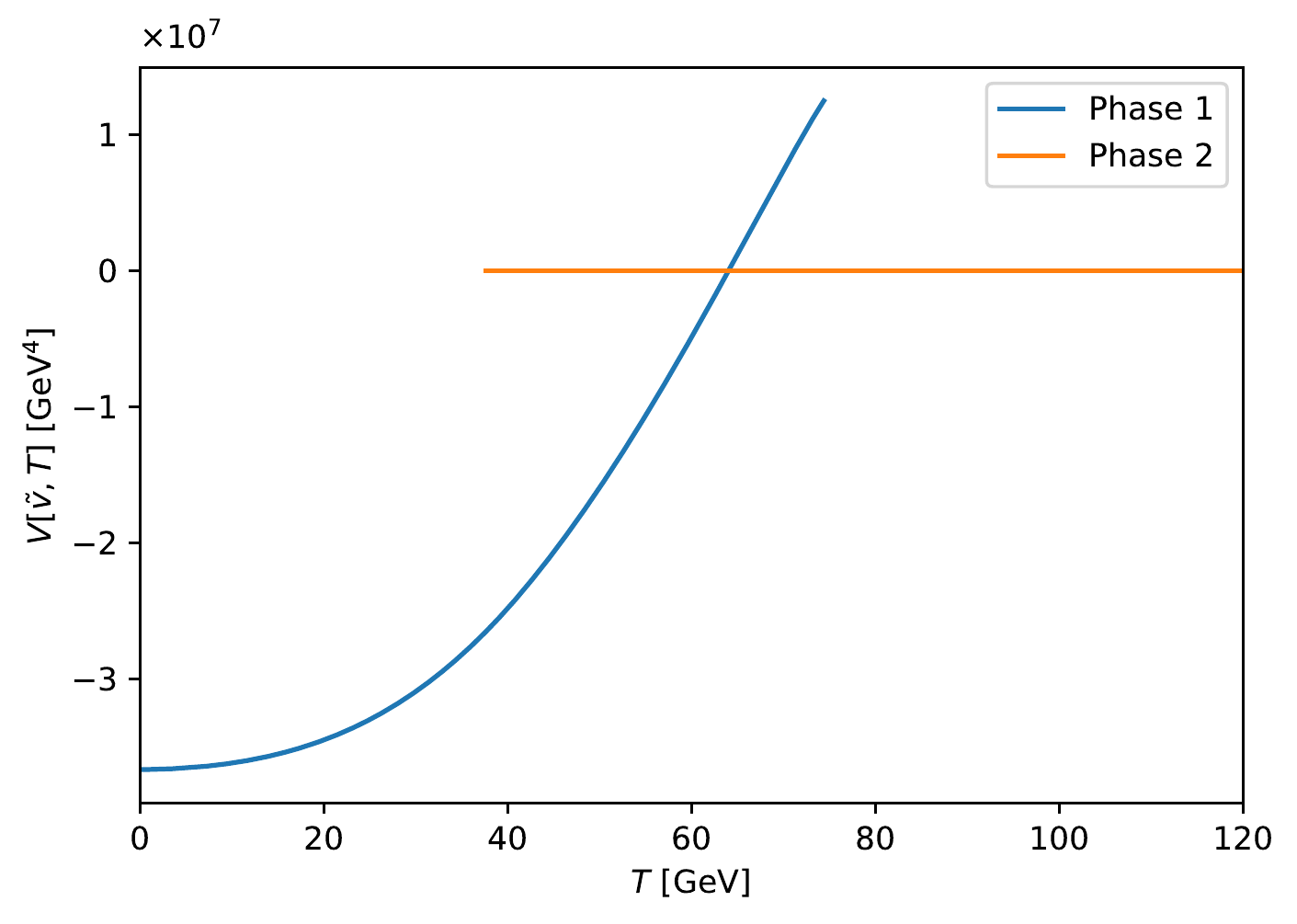}
	\caption{Evolution of $V[\tilde{v},T]$ for one-step phase transition,$\tilde{v} = \sqrt{\tilde{v}_1^2 + \tilde{v}_2^2 + \tilde{v}_{CP}^2 + \tilde{v}_{CB}^2}$.}\label{Veff1}
\end{figure}

\begin{figure}
	\centering
	\includegraphics[scale=0.8]{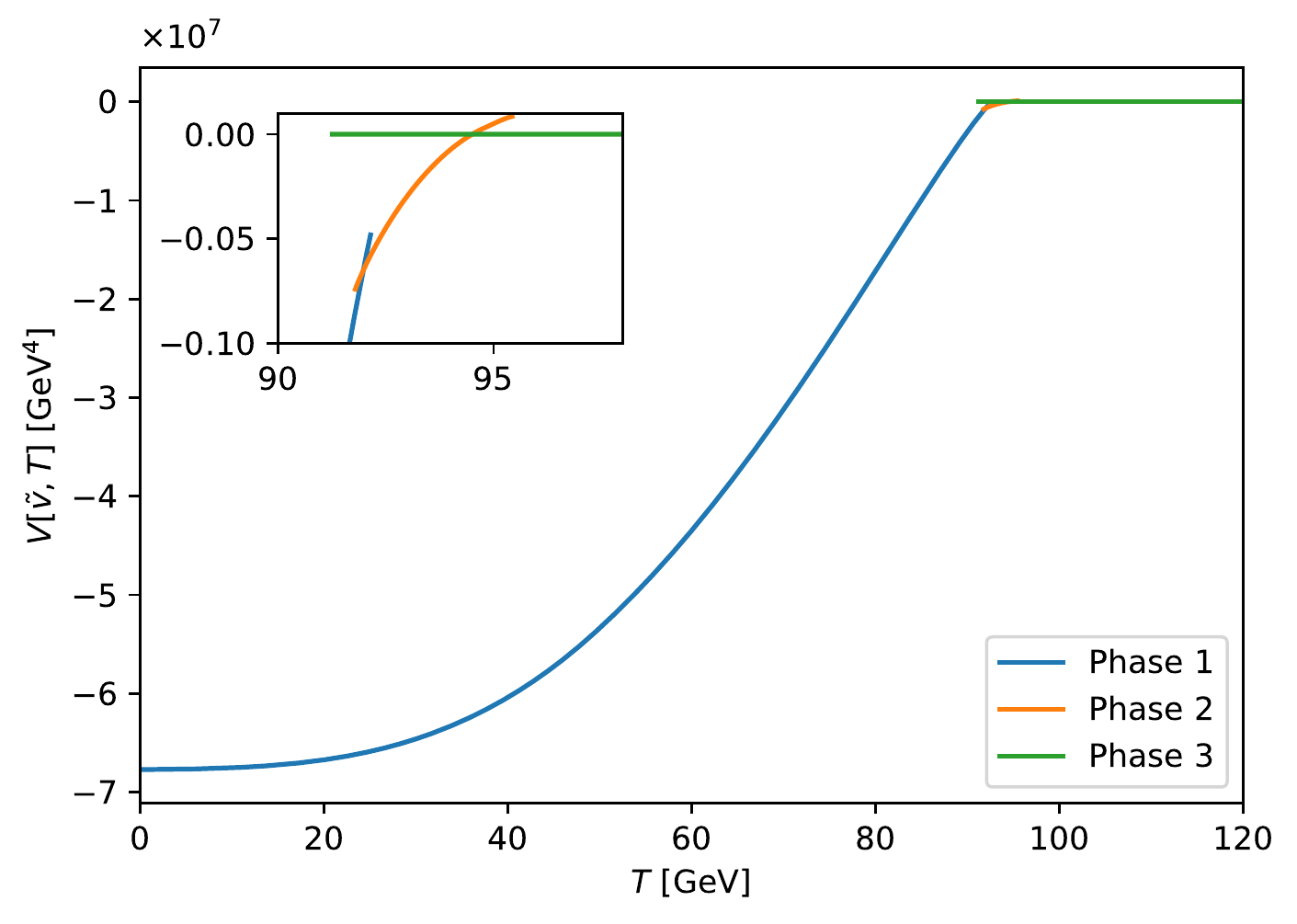}
	\caption{Evolution of $V[\tilde{v},T]$ for two-step phase transition.}\label{Veff2}
\end{figure}

\begin{figure}
	\centering
	\includegraphics[scale=0.8]{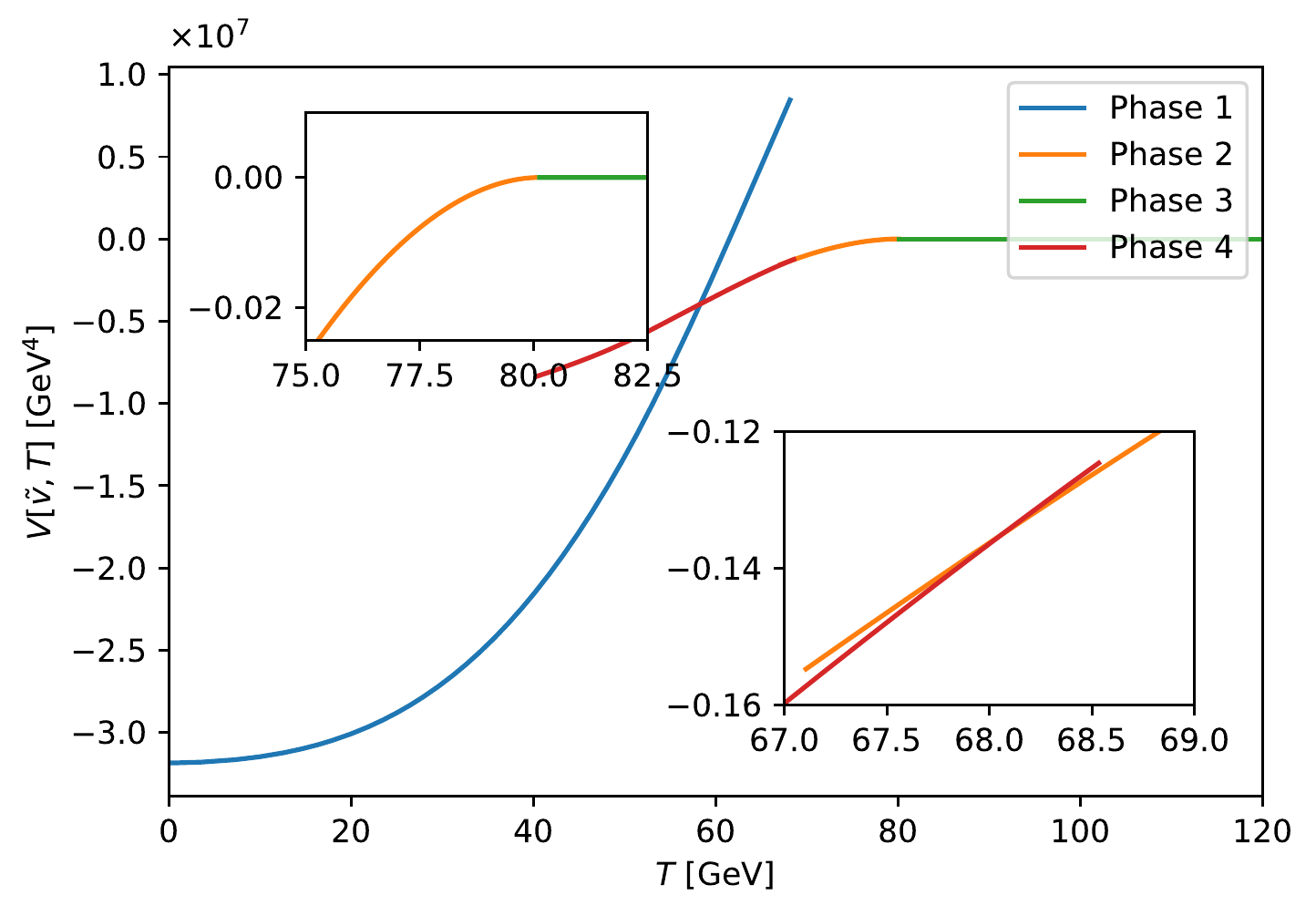}
	\caption{Evolution of $V[\tilde{v},T]$ for three-step phase transition.}\label{Veff3}
\end{figure}

As mentioned above, we assume CP-violating VEV $\tilde{v}_{CP}$  can get non-zero value at finite temperature and disappear at zero temperature.
To verify our assumption, we do the numerical consistent check for the different phase transition patterns,
and show the evolution of CP-violating VEV together with $\tilde{v}_1$ an $\tilde{v}_2$ in Fig.~\ref{wx1}, Fig.~\ref{wx3}, and Fig.~\ref{wx4}, respectively.
For example, Fig.~\ref{wx1} depicts the one-step FOPT pattern, and it shows that the CP-violating VEV $\tilde{v}_{CP}$  increases with temperature, when it is below critical temperature.
And when the temperature decreases, the  CP-violating VEV $\tilde{v}_{CP}$ gradually evolves to zero.
This is consistent with our assumption.
For other phase transition patterns, they are also consistent.
As for the charge-breaking VEV, it also numerically shows the similar behavior except the VEV value is much smaller compared to CP-violating case.
The extra CP-violating source at finite temperature may provide enough CP violation for successful
EW baryogenesis. And this extra CP-violating source evolves to zero at zero temperature to avoid the strong constraints from EDM data.
Furthermore, we also show the value of the effective potential at the vacua as a function of temperature in Fig.~\ref{Veff1}, Fig.~\ref{Veff2}, and Fig.~\ref{Veff3}.
For different phase transition patterns, the value of effective potential at vacua shows very distinctive behavior.
Taking Fig.~\ref{Veff3} as an example, we can see the evolution behavior of a SOPT (from phase-3 to phase-2), is different from the FOPT processes (phase-2 to phase-4 and phase-4 to phase-1).

\subsection{Supercooling case}
The benchmark sets $BP_3$, $BP_4$, $BP_7$ and $BP_8$ can produce supercooling pattern of FOPT, where $\alpha >1$.
In this case, according to the definition of $\alpha$, the PT latent heat density dominates the plasma energy density.
In the previous figures, we show the GW signals for the $\alpha \ll 1$ since both the bubble dynamics and GW spectra are well studied.
However, for $\alpha >1$ case ($BP_3$, $BP_4$, $BP_7$ and $BP_8$ ), it is still unclear and under investigating, such as Ref.~\cite{Jinno:2019jhi}.
For a strong supercooling, bubbles become very thin and relativistic.
In this case, the bubble wall velocity $v_b$ quickly approaches the speed of light.
This is the so-called runaway bubbles in vacuum~\cite{Caprini:2015zlo}, which means phase transitions occur in a vacuum-dominated epoch.
In principle, these benchmark sets can trigger even stronger GW signals.
However, it is still controversial~\cite{Ellis:2018mja}.
Therefore, we leave the precise study of the GW spectrum for the supercooling case in our future work.
As for the implication from this C2HDM, from numerical calculations, we find that supercooling favors relatively large coupling constants.
And in some narrow parameter spaces, the nucleation temperature decreases as the mass hierarchy of the two neutral Higgs bosons decreases. More reliable results rely on further lattice simulations.

\section{Conclusion}
We have studied the phase transition dynamics in detail with the existence of dynamical CP-violation at finite-temperature in the complex two-Higgs doublet Model. Various phase transition patterns have been investigated, including multi-step phase transition and supercooling case in this scenario.
The dynamical CP-violation can not only provide a possible cosmological origin of CP-violation source, but also make the phase transition dynamics more abundant.
The corresponding GW signals in synergy with collider signals have also been discussed, which can be used to make complementary
test on this scenario and further unravel the underlying phase transition dynamics or different patterns in the early universe.
The detailed study on the realization of the EW baryogenesis and GWs from supercooling are left for our future work.

\begin{acknowledgments}
XW would like to thank Phillipp Basler for correspondence regarding the \emph{BSMPT} package, as well as Carroll L. Wainwright for useful discussions of the \emph{cosmoTranstions} package. FPH deeply appreciates Eibun Senaha's various discussion.
We appreciate the valuable comments from the anonymous referee.
XW and XMZ are supported in part by  the Ministry of Science and Technology of China (2016YFE0104700), the National Natural Science Foundation of China (Grant NO. 11653001), the CAS pilot B project (XDB23020000).
FPH is supported in part by the McDonnell Center for the Space Sciences.
\end{acknowledgments}

\begin{appendices}
	
\section{APPENDIX A:Temperature correction of mass in C2HDM}\label{A}
There are contributions to the ring diagrams from the gauge bosons and Higgs bosons. We need to calculate the self-energy of the gauge bosons and Higgs bosons in the IR limit. First, we consider the self-energy of the Higgs bosons. The Higgs self-energy can be derived from the propagator of the Higgs bosons with Higgs bosons, gauge bosons, and top quark loops. We work in the original basis, where the relevant fields are $\phi_i\equiv\{\rho_1,\eta_1,\rho_2,\eta_2,\zeta_1,\psi_1,\zeta_2,\psi_2\}$, then the contributions to the Higgs self-energy from the Higgs bosons are \cite{Basler:2016obg,Basler:2017uxn,Cline:1996mga,Cline:2011mm}
\begin{equation}
{\Pi}_{\phi_i\phi_i}^S = \frac{T^2}{24}(6\lambda_1+4\lambda_3+ 2\lambda_4)  \quad \phi_i = \{\rho_1,\eta_1,\zeta_1,\psi_1\},
\end{equation}
\begin{equation}
{\Pi}_{\phi_i\phi_i}^S = \frac{T^2}{24}(6\lambda_2 + 4\lambda_3+ 2\lambda_4) \quad \phi_i = \{\rho_2,\eta_2,\zeta_2,\psi_2\}.
\end{equation}
The contribution comes from the gauge bosons is \cite{Basler:2016obg,Basler:2017uxn,Cline:1996mga,Cline:2011mm}
\begin{equation}
{\Pi}_{\phi_i\phi_i}^{GB} = \frac{T^2}{16}(3g^2+ {g^\prime}^2).
\end{equation}
The contribution from top-quark loop is \cite{Basler:2016obg,Basler:2017uxn,Cline:1996mga,Cline:2011mm}
\begin{equation}
{\Pi}_{\phi_i\phi_i}^F = \frac{T^2}{4}y_t^2\quad \phi_i = \{\rho_2,\eta_2,\zeta_2,\psi_2\}.
\end{equation}
Thus, the total contributions to the Higgs boson self-energy in the C2HDM are \cite{Basler:2016obg,Basler:2017uxn,Cline:1996mga,Cline:2011mm}
\begin{equation}
\Pi_{\phi_i\phi_i}^1 = \frac{T^2}{48}(12\lambda_1 + 8\lambda_3 + 4\lambda_4 +3(3g^2 + {g^\prime}^2)) \quad \phi_i = \{\rho_1,\eta_1,\zeta_1,\psi_1\},
\end{equation}
\begin{equation}
\Pi_{\phi_i\phi_i}^2 = \frac{T^2}{48}(12\lambda_2 + 8\lambda_3 + 4\lambda_4 + 3(3g^2 + {g^\prime}^2) + 12y_t^2) \quad \phi_i = \{\rho_2,\eta_2,\zeta_2,\psi_2\}.
\end{equation}

Next, we calculate the self-energy of gauge bosons. There are two relevant fields in original basis $W_{\mu}^a,B_{\mu}$.
Then the contributions to the gauge bosons self-energy come from the gauge bosons, Higgs bosons, and top quark, respectively.
Hence, the total self-energy for the gauge bosons in the C2HDM are \cite{Basler:2016obg,Basler:2017uxn,Cline:1996mga,Cline:2011mm}
\begin{equation}
{\Pi}_{W^aW^a} = 2g^2T^2, \notag
\end{equation}
\begin{equation}
\Pi_{BB} = 2{g^\prime}^2T^2.
\end{equation}

\section{APPENDIX B:Field dependent mass matrix elements of C2HDM}\label{B}
Since we introduce a charge-breaking VEV, the mass matrix of gauge bosons and Higgs bosons in the original basis are fully mixed. We can not give the analytic form of the field-dependent mass for each physical particle. Instead, we derive the mass matrix in the original basis, and then numerically calculate the eigenvalues which are the physical masses of the particles. The field-dependent mass matrix elements of the gauge bosons in the original basis can be written as \cite{Basler:2016obg,Cline:1996mga,Cline:2011mm}
\begin{equation}
m_{11}^G = m_{22}^G = m_{33}^G = \frac{1}{4}g^2(\tilde{v}_1^2 + \tilde{v}_2^2 + \tilde{v}_{CP}^2 + \tilde{v}_{CB}^2), \notag
\end{equation}
\begin{equation}
m_{44}^G = \frac{1}{4}{g^\prime}^2(\tilde{v}_1^2 + \tilde{v}_2^2 + \tilde{v}_{CP}^2 + \tilde{v}_{CB}^2), \notag
\end{equation}
\begin{equation}
m_{14}^G = \frac{1}{2}gg^\prime\tilde{v}_{2}\tilde{v}_{CB}, \notag
\end{equation}
\begin{equation}
m_{24}^G = \frac{1}{2}gg^\prime\tilde{v}_{CP}\tilde{v}_{CB}, \notag
\end{equation}
\begin{equation}
m_{34}^G = -\frac{1}{4}gg^\prime(\tilde{v}_1^2 + \tilde{v}_2^2 + \tilde{v}_{CP}^2 - \tilde{v}_{CB}^2).
\end{equation}
The mass matrix is
\begin{equation}
M_{GB} = \left(\begin{array}{cccc}
m_{11}^G&0&0&m_{14}^G \\
0&m_{22}^G&0& m_{24}^G \\
0&0&m_{33}^G& m_{34}^G\\
m_{14}^G&m_{24}^G& m_{34}^G& m_{44}^G
\end{array}\right).
\end{equation}
For the longitudinal components of the gauge bosons, we need to consider the Debye corrected masses, which are the eigenvalues of
\begin{equation}
\overline{M}_{GB} = M_{GB} + \text{diag}(\Pi_{W^aW^a},\Pi_{W^aW^a},\Pi_{W^aW^a}, \Pi_{BB}).
\end{equation}
The mass matrix elements of Higgs bosons in the original basis can be expressed as \cite{Basler:2016obg,Cline:1996mga,Cline:2011mm}
\begin{equation}
M_{11}=m_{11}^2 + \frac{1}{2}\lambda_1\tilde{v}_{1}^2 + \frac{1}{2}\lambda_3(\tilde{v}_{1}^2 + \tilde{v}_{CP}^2) + \frac{1}{2}\lambda_{345}\tilde{v}_{CB}^2, \notag\\
\end{equation}
\begin{equation}
M_{22}= m_{11}^2 + \frac{1}{2}\lambda_1\tilde{v}_{1}^2 + \frac{1}{2}\lambda_3(\tilde{v}_{1}^2 + \tilde{v}_{CP}^2) + \frac{1}{2}\bar{\lambda}_{345}\tilde{v}_{CB}^2, \notag
\end{equation}
\begin{equation}
M_{33} = m_{22}^2 + \frac{1}{2}\lambda_2(\tilde{v}_{2}^2 + \tilde{v}_{CP}^2 + 3\tilde{v}_{CB}^2) + \frac{1}{2}\lambda_3\tilde{v}_{1}^2, \notag
\end{equation}
\begin{equation}
M_{44} =  m_{22}^2 + \frac{1}{2}\lambda_2(\tilde{v}_{2}^2 + \tilde{v}_{CP}^2 + \tilde{v}_{CB}^2) + \frac{1}{2}\lambda_3\tilde{v}_{1}^2, \notag
\end{equation}
\begin{equation}
M_{55} = m_{11}^2 + \frac{3}{2}\lambda_3\tilde{v}_{1}^2 + \frac{1}{2}\lambda_{345}\tilde{v}_{2}^2 + \frac{1}{2}\bar{\lambda}_{345}\tilde{v}_{CP}^2 + \frac{1}{2}\lambda_3\tilde{v}_{CB}^2 - Im(\lambda_5)\tilde{v}_{2}\tilde{v}_{CP}, \notag
\end{equation}
\begin{equation}
M_{66} =  m_{11}^2 + \frac{1}{2}\lambda_3\tilde{v}_{1}^2 + \frac{1}{2}\lambda_{345}\tilde{v}_{CP}^2 + \frac{1}{2}\bar{\lambda}_{345}\tilde{v}_{2}^2 + \frac{1}{2}\lambda_3\tilde{v}_{CB}^2 + Im(\lambda_5)\tilde{v}_{2}\tilde{v}_{CP}, \notag
\end{equation}
\begin{equation}
M_{77} = m_{22}^2 + \frac{1}{2}\lambda_2(3\tilde{v}_{2}^2 + \tilde{v}_{CP}^2 + \tilde{v}_{CB}^2) + \frac{1}{2}\lambda_{345}\tilde{v}_{1}^2, \notag
\end{equation}
\begin{equation}
M_{88} = m_{22}^2 + \frac{1}{2}\lambda_2(\tilde{v}_{2}^2 + 3\tilde{v}_{CP}^2 + \tilde{v}_{CB}^2) + \frac{1}{2}\bar{\lambda}_{345}\tilde{v}_{1}^2, \notag
\end{equation}
\begin{equation}
M_{12} = \frac{1}{2}Im(\lambda_5)\tilde{v}_{CB}^2, \notag
\end{equation}
\begin{equation}
M_{13} = - Re(m_{12}^2) - \frac{1}{2}Im(\lambda_5)\tilde{v}_{1}\tilde{v}_{CP} + \frac{1}{2}(Re(\lambda_5) + \lambda_4)\tilde{v}_{1}\tilde{v}_{CP}, \notag
\end{equation}
\begin{equation}
M_{14} = Im(m_{12}^2) - \frac{1}{2}Im(\lambda_5)\tilde{v}_1\tilde{v}_2  + \frac{1}{2}(\lambda_4 - Re(\lambda_5))\tilde{v}_{1}\tilde{v}_{CP}, \notag
\end{equation}
\begin{equation}
M_{15} = \frac{1}{2}\left[(\lambda_4 + Re(\lambda_5))\tilde{v}_2\tilde{v}_{CB} - Im(\lambda_5)\tilde{v}_{CP}\tilde{v}_{CB}\right], \notag
\end{equation}
\begin{equation}
M_{16} = \frac{1}{2}\left[Im(\lambda_5)\tilde{v}_{2}\tilde{v}_{CB} + (\lambda_4 + Re(\lambda_5))\tilde{v}_{CP}\tilde{v}_{CB}\right], \notag
\end{equation}
\begin{equation}
M_{17} = \frac{1}{2}(\lambda_4 + Re(\lambda_5))\tilde{v}_1\tilde{v}_{CB}, \notag
\end{equation}
\begin{equation}
M_{18} = -\frac{1}{2}Im(\lambda_5)\tilde{v}_1\tilde{v}_{CB}, \notag
\end{equation}
\begin{equation}
M_{23} = -Im(m_{12}^2) + \frac{1}{2}Im(\lambda_5)\tilde{v}_1\tilde{v}_2 +\frac{1}{2}(Re(\lambda_5) - \lambda_4)\tilde{v}_1\tilde{v}_{CP}, \notag
\end{equation}
\begin{equation}
M_{24} = -Re(m_{12}^2) + \frac{1}{2}(\lambda_4 + Re(\lambda_5))\tilde{v}_{1}\tilde{v}_{2} - \frac{1}{2}Im(\lambda_5)\tilde{v}_{1}\tilde{v}_{CP}, \notag
\end{equation}
\begin{equation}
M_{25} = \frac{1}{2}\left[Im(\lambda_5)\tilde{v}_{2}\tilde{v}_{CB} - (\lambda_4 - Re(\lambda_5))\tilde{v}_{CP}\tilde{v}_{CB} \right], \notag
\end{equation}
\begin{equation}
M_{26} = \frac{1}{2}\left[(\lambda_4 - Re(\lambda_5))\tilde{v}_{2}\tilde{v}_{CB} + Im(\lambda_5)\tilde{v}_{CP}\tilde{v}_{CB}\right], \notag
\end{equation}
\begin{equation}
M_{27} = \frac{1}{2}Im(\lambda_5)\tilde{v}_{1}\tilde{v}_{CB}, \notag
\end{equation}
\begin{equation}
M_{28} = -\frac{1}{2}(\lambda_4 - Re(\lambda_5))\tilde{v}_1\tilde{v}_{CB}, \notag
\end{equation}
\begin{equation}
M_{35} = \lambda_3\tilde{v}_1\tilde{v}_{CB}, \notag
\end{equation}
\begin{equation}
M_{37} = \lambda_2\tilde{v}_2\tilde{v}_{CB}, \notag
\end{equation}
\begin{equation}
M_{38} = \lambda_2\tilde{v}_{CP}\tilde{v}_{CB}, \notag
\end{equation}
\begin{equation}
M_{56} = Re(\lambda_5)\tilde{v}_2\tilde{v}_{CP} + \frac{1}{2}Im(\lambda_5)(\tilde{v}_{2}^2 - \tilde{v}_{CP}^2), \notag
\end{equation}
\begin{equation}
M_{57} = - Re(m_{12}^2) + \lambda_{345}\tilde{v}_{1}\tilde{v}_2 - Im(\lambda_5)\tilde{v}_1\tilde{v}_{CP}, \notag
\end{equation}
\begin{equation}
M_{58} = Im(m_{12}^2) - Im(\lambda_5)\tilde{v}_1\tilde{v}_2 + \bar{\lambda}_{345}\tilde{v}_{1}\tilde{v}_{CP}, \notag
\end{equation}
\begin{equation}
M_{67} = -Im(m_{12}^2) + Re(\lambda_5)\tilde{v}_{1}\tilde{v}_{CP} + Im(\lambda_5)\tilde{v}_{1}\tilde{v}_{2}, \notag
\end{equation}
\begin{equation}
M_{68} = - Re(m_{12}^2) + Re(\lambda_5)\tilde{v}_1\tilde{v}_2 - Im(\lambda_5)\tilde{v}_1\tilde{v}_{CP}, \notag
\end{equation}
\begin{equation}
M_{78} = -\frac{1}{2}Im(\lambda_5)\tilde{v}_{1}^2 + \lambda_2\tilde{v}_2\tilde{v}_{CP},
\end{equation}
where
\begin{gather}
\lambda_{345} = \lambda_3 + \lambda_4 + Re(\lambda_5), \notag\\
\bar{\lambda}_{345} = \lambda_3 + \lambda_4 - Re(\lambda_5).
\end{gather}
The mass matrix is
\begin{equation}
M_S = \left(\begin{array}{cccccccc}
M_{11}&M_{12}&M_{13}&M_{14}&M_{15}&M_{16}&M_{17}&M_{18} \\
M_{12}&M_{22}&M_{23}&M_{24}&M_{25}&M_{26}&M_{27}&M_{28}\\
M_{13}&M_{23}&M_{33}&0&M_{35}&0&M_{37}&M_{38}\\
M_{14}&M_{24}&0&M_{44}&0&0&0&0\\
M_{15}&M_{25}&M_{35}&0&M_{55}&M_{56}&M_{57}&M_{58}\\
M_{16}&M_{26}&0&0&M_{56}&M_{66}&M_{67}&M_{68} \\
M_{17}&M_{27}&M_{37}&0&M_{57}&M_{67}&M_{77}&M_{78}\\
M_{18}&M_{28}&M_{38}&0&M_{58}&M_{68}&M_{78}&M_{88}
\end{array}\right).
\end{equation}
The Debye corrected mass of the scalar bosons are given as the eigenvalues of
\begin{equation}
\overline{M}_S = M_S + \text{diag}(\Pi_{\phi_i\phi_i}^1, \Pi_{\phi_i\phi_i}^1, \Pi_{\phi_i\phi_i}^2, \Pi_{\phi_i\phi_i}^2,\Pi_{\phi_i\phi_i}^1, \Pi_{\phi_i\phi_i}^1, \Pi_{\phi_i\phi_i}^2, \Pi_{\phi_i\phi_i}^2)\,\,.
\end{equation}
Since we just consider the top quark in our work, the field dependent mass of top quark can be easily derived as
\begin{equation}
m_t^2 = \frac{1}{2}y_t^2(\tilde{v}_{2}^2 + \tilde{v}_{CP}^2)\,\,.
\end{equation}

\end{appendices}

\end{document}